\documentclass[%
 reprint,
 superscriptaddress,
 amsmath,amssymb,
 aps,
]{revtex4-2}

\usepackage{graphicx}% Include figure files
\usepackage{dcolumn}% Align table columns on decimal point
\usepackage{bm}% bold math
\usepackage[version=4]{mhchem}

\usepackage{subfig} 
\usepackage{verbatim}
\usepackage{graphics,setspace,float,enumitem}
\usepackage{commath,mathrsfs} % amsbsy, mathabx
\usepackage[framemethod=TikZ]{mdframed}
\usepackage[noabbrev,capitalize]{cleveref}
\usepackage{multirow}

\creflabelformat{equation}{#2\textup{#1}#3}

\newcommand{\spso}{\texttt{SpinPSO}}

\newcommand{\Fsf}[1]{#1\textsubscript{SF}}
\newcommand{\FpUJ}[1]{#1+$U$+$J$}
\newcommand{\FsfUJ}[1]{\FpUJ{\Fsf{#1}}}

\newcommand{\SpinPSOtoSFcaption}[1]{Computed ground-state magnetic configurations for \ce{#1} obtained from \spso{} with PBE for multiple randomly initialized \spso{} trajectories. On the left are the minimum energy configurations computed from \spso{} with PBE, which are used as input structures for \Fsf{PBE}, with ground-state structures shown on the right.}

\begin{document}

\preprint{APS/123-QED}

\title{\spso{}: A computational optimization workflow for identifying noncollinear magnetic ground-states from first-principles}% Force line breaks with \\

\author{Guy C. Moore}
\affiliation{Department of Materials Science and Engineering, University of California Berkeley, Berkeley, CA 94720, USA}
\affiliation{Materials Science Division, Lawrence Berkeley National Laboratory, Berkeley, CA 94720, USA}

\author{Matthew K. Horton}
\affiliation{Department of Materials Science and Engineering, University of California Berkeley, Berkeley, CA 94720, USA}
\affiliation{Materials Science Division, Lawrence Berkeley National Laboratory, Berkeley, CA 94720, USA}

\author{Kristin A. Persson}
\affiliation{Department of Materials Science and Engineering, University of California Berkeley, Berkeley, CA 94720, USA}
\affiliation{Molecular Foundry, Lawrence Berkeley National Laboratory, Berkeley, CA 94720, USA}


\date{\today}% It is always \today, today,

\begin{abstract}

We propose and implement a novel hybrid meta-heuristic optimization algorithm for the identification of non-collinear global ground-states in magnetic systems. The inputs to this optimization scheme are directly from non-collinear density functional theory (DFT), and the workflow is implemented in \texttt{atomate}, making it suitable to run on high-performance computing architectures. The hybrid algorithm provides a seamless theoretical extension of particle swarm optimization (PSO) algorithms to continuous $\mathcal S^2$ spins, giving it the name \spso{}. The hybrid nature of the algorithm stems from setting the dynamics of individual spins to be governed by physically-motivated atomistic spin dynamics. Using this algorithm, we are able to achieve convergence to experimentally resolved magnetic ground-states for a set of diverse test case materials that exhibit exotic spin textures.

\begin{comment}
\begin{description}
\item[Usage]
Secondary publications and information retrieval purposes.
\item[Structure]
You may use the \texttt{description} environment to structure your abstract;
use the optional argument of the \verb+\item+ command to give the category of each item. 
\end{description}
\end{comment}
\end{abstract}

\keywords{}  % Use showkeys class option if keyword display desired

\maketitle

\section{Introduction}

Identifying non-collinear magnetic ground-states is a complex and challenging problem in materials physics. Several different phenomena can contribute to energy preferred non-collinear spin configurations, from spin-orbit coupling (SOC) \cite{parkStructureSpinDynamics2014a}, to magnetic frustration, both geometric and other \cite{lyonsMethodDeterminingGroundState1960, rossNanosizedHelicalMagnetic2015}. 
Previous computational and theoretical studies have had success in predicting non-collinear magnetic orderings using advanced symmetry-based models informed by first-principles for collinear \cite{hortonHighthroughputPredictionGroundstate2019a, freyHighthroughputSearchMagnetic2020a} and noncollinear \cite{huebschBenchmarkInitioPrediction2021, suzukiMultipoleExpansionMagnetic2019} magnetic systems. The incorporation of symmetry has benefits from the standpoint of reduced computational cost \cite{huebschBenchmarkInitioPrediction2021, suzukiMultipoleExpansionMagnetic2019}. However, these approaches will not be suitable for systems of lower or broken crystal symmetry, such as in systems with interfaces or other defects, or amorphous magnets. Although we do not study these systems here, our methodology could be applicable in these instances, which we hope is a topic of future exploration.

In our algorithm, we modify the  guaranteed convergence particle swarm optimization (GCPSO) algorithm developed by Ref.~\onlinecite{vandenberghNewLocallyConvergent2002}~\&~\onlinecite{peerUsingNeighbourhoodsGuaranteed2003} by incorporating a variation of the well-known atomistic Landau Lifshitz Gilbert (LLG) equation \cite{ASD-FoundAndApp, cortes-ortunoThermalStabilityTopological2017} into the PSO methodology. For this reason, the algorithm has been titled ``\spso{}." This modification allows the facile extension to dynamics on $\mathcal S^2$, compared to the important work of Ref.~\onlinecite{payneFireflyAlgorithmApplied2018}, in which the authors necessitate the definition of a distance metric between spins. In our implementation, we call on the ubiquitous and physically motivated LLG equation, which is known to describe the magnetization dynamics at multiple length scales \cite{ASD-FoundAndApp, lakshmananFascinatingWorldLandau2011, BERTOTTI199873}. 
% If we were to be pedantic, we employ the Landau Lifshitz (LL) equation, in its original form \cite{}. The LL and LLG equation are often conflated, but for good reason, because in the case of spins with constant magnitude, the two can be shown to be equivalent \cite{}.

Agent-based meta-heuristic schemes often have improved global convergence \cite{vandenberghNewLocallyConvergent2002}, in the sense that a large initial scattering of agents on the potential energy landscape reduces the possiblility of being trapped in a local minimum, compared to gradient descent based optimization schemes. Ref.~\onlinecite{payneFireflyAlgorithmApplied2018} has provided inspiration to apply agent-based optimization towards optimizing noncollinear magnetic orderings. Payne et al. use a firefly optimization algorithm - the cousin of PSO. This is possibly because firefly has been shown to outperform PSO for noisy objective functions \cite{palComparativeStudyFirefly2012}. However, density functional theory (DFT) calculations, derived from variational principles, should only have noise on the order of the energy convergence tolerance. Therefore, in this study, we use PSO because in many cases it converges twice as rapidly than the firefly algorithm  \cite{bhushanParticleSwarmOptimization2013, jonesComparisonFireflyAlgorithm2011a}.

In the formulation that we found to be the most successful, \spso{} is further distinct from conventional GCPSO in the sense that we have included gradient information in the optimization scheme. This is not a conventional practice in most implementations of PSO, which prides itself on being ``gradient-free'' \cite{vandenberghNewLocallyConvergent2002, peerUsingNeighbourhoodsGuaranteed2003}. However, there is no rule that states that GCPSO cannot incorporate gradient information. After all, GCPSO was motivated on the grounds that PSO-type optimization schemes perform well when identifying global valleys of the potential energy surface (PES) \cite{vandenberghNewLocallyConvergent2002}. However, the order of convergence can become sluggish - and even stuck in some instances - within the same convex hull of the global minimum \cite{vandenberghNewLocallyConvergent2002, peerUsingNeighbourhoodsGuaranteed2003}. Therefore, within the GCPSO formalism, a random noise term is added to the velocity of the agent with the ``best'' position on the PES \cite{vandenberghNewLocallyConvergent2002}. This perturbation allows further exploration of the local energy landscape, in order to ``guarantee convergence.''

Therefore, inspired by gradient-descent type optimization strategies, we inform the ``best agent'' dynamics with gradient information, as opposed to a fluctuating noise term. While this doesn't appear to be common knowledge within the non-collinear DFT community, information on the gradient, or a local effective magnetic fields, can be obtained from constraints on local magnetic moments, as implemented in VASP \cite{maConstrainedDensityFunctional2015, mooreHighthroughputDeterminationHubbard2022a}. The details of how we compute these local effective fields, and incorporate them into the workflow, are included in Section \ref{sec:GradInfo}. We also explore how including gradient information improves convergence speed by orders of magnitude. This possibly explains why we are able to achieve much higher rates of convergence compared with other agent-based optimization strategies for computing the noncollinear magnetic ground-states \cite{payneFireflyAlgorithmApplied2018}.

\section{Methods}

\subsection{Code availability}
\label{sec:CodeAvail}

The \spso{} optimizer is implemented in both the \texttt{pymatgen} \cite{pymatgen-github} and \texttt{atomate} \cite{atomate-paper, atomate-github} codebases, which we provide in Ref.'s~\onlinecite{pymatgen-spinpso-github}~\&~\onlinecite{atomate-spinpso-github}. These include the default input settings for the Vienna Ab-Initio Simulation Program (VASP) code \cite{VASPpaper}. The Hubbard $U$ and Hund $J$ values used in this study are the mean values reported in Ref.~\onlinecite{mooreHighthroughputDeterminationHubbard2022a}.

\subsection{Mathematical foundation of \spso{}}

This optimization strategy is comprised of a swarm of agents, in which each agent, or particle, corresponds to a different magnetic configuration. Each agent's position on the potential energy landscape is evolved in a fictitious ``time" $\tau$ according to the atomistic LLG equation
\begin{align}
    \frac{\Delta \bm{s}_{i,j}}{\Delta \tau} 
    &= \frac{\bm{s}_{i,j}^{n+1} - \bm{s}_{i,j}^{n}}{\Delta \tau} \nonumber \\
    &= - \gamma \bm{s}_{i,j}^{n} \times \bm{h}_{i,j}^{n} - \alpha \bm{s}_{i,j}^{n} \times  \left( \bm{s}_{i,j}^{n} \times \bm{h}_{i,j}^{n} \right)
\end{align}
Each ``agent'' $i$ corresponds to a spin configuration $\left\{\bm s_{i,1}, ..., \bm s_{i,n}\right\}$, where each atomic moment is normalized within $\mathbb R^3$, s.t. $s_{i,j} \in \mathcal S^2$. Each spin is evolved in $\tau$ according to a forward-difference Landau Lifshitz Gilbert (LLG) equation at interval $\Delta \tau$ indexed by time-step $n$. To ensure that the norms of the spins remain constant ($|\bm s| = 1$), we perform a simple re-scaling of the individual spins after each time step. We do not employ a more advanced time-stepping routine (e.g. mid-point method), because we are currently more concerned with convergence to the minimum energy ground-state than accurately describing the kinetics of the spins themselves.

The two terms on the right-hand side of the LLG equation can be intuited as follows. The first, single cross product term governs the precession of the moments about the effective field. The second term, the triple product, promotes a damping motion of the moments, resulting in a tendency to align with the on-site effective field, $\bm{h}$. These terms have physical meaning; after all, the LLG equation can be derived from the quantum mechanical Heisenberg time evolution equation for spin operators \cite{ASD-FoundAndApp} -- at least without damping ($\alpha = 0$). The LLG equation has demonstrated effectiveness at describing magnetization dynamics at multiple length scales \cite{lakshmananFascinatingWorldLandau2011}, and its use is often motivated on phenomenological grounds for this reason \cite{BERTOTTI199873}. 

In the context of our algorithm, the second, damping term is the most important, because at convergence/equilibrium the spins should eventually align with the effective field, which contains information on the swarm's best position in the configuration space. The first term on the right-hand side induces a precession about best state in the pure particle swarm optimization scheme, which allows for slight curvature in the path towards convergence, which should in practice increase exploration of the potential energy landscape near to the identified lower energy state. We don't include inertial terms in our optimization scheme, as is common practice in particle swarm optimization dynamical schemes. Therefore, we include this precession term instead. For stability and convergence, we choose the coefficient of the second, $\alpha$, to be significantly higher than the first, $\gamma$. We found a good rule of thumb to be at least a factor of three difference.

\subsection{Swarm and gradient informed local fields}
\label{sec:GradInfo}

The effective field on individual moments, $\bm{h}_{i,j}^{n}$, is informed by particle swarm optimization input quantities from the swarm of agents,
\begin{align}
    \bm{l}_{i,j}^{n} &= a_c \sigma_c \bm{\tilde s}_{i,j}^{n} + a_s \sigma_s {\bm{\hat s}}_{j}^{n} \nonumber \\
    \bm{h}_{i,j}^{n} &= \frac{\bm{l}_{i,j}^{n}}{| \bm{l}_{i,j}^{n} |}
\end{align}
where $(1 - \sigma_c) \sim \mathcal U\left(0,1\right)$ and $(1 - \sigma_s) \sim \mathcal U\left(0,1\right)$. The local fields are normalized stochastic linear combination of $\bm{\tilde s}_{i,j}^{n}$ and $\bm{\hat s}_{j}^{n}$, which are the ``personal best'' and ``swarm best'' spin configurations, respectfully, consistent with the PSO formalism. Particle swarm optimization algorithms are inherently stochastic, so in an analogy to their Cartesian counterparts, we introduced these random uniform weighting of the cognition term, which corresponds to the historical best position of the corresponding agent over its trajectory directory. In addition, we include the social term characteristic of particle swarm optimization schemes, which stores the best position compared to each time step of all agents over time.

To remain consistent with the guaranteed convergence particle swarm optimization (GCPSO) scheme developed by van den Bergh and Engelbrecht \cite{vandenberghNewLocallyConvergent2002}, we define an alternative time evolution equation to the ``best" agent at any particular time. We replace the random perturbations of the ``best'' agent evolution in the guaranteed convergence PSO (GCPSO) \cite{vandenberghNewLocallyConvergent2002} with gradient information from local moment constraints implemented in VASP, indicated by $\left(\bm{h}_{\text{eff}}\right)_{i,j}^{n}$.
\begin{align}
    \left(\bm{h}_{\text{eff}}\right)_{i,j}^{n} &= - \frac{\partial E}{\partial \bm{s}_{i,j}^{n}} \nonumber \\
    \bm{\hat l}_{i,j}^{n} &= \rho \frac{\left(\bm{h}_{\text{eff}}\right)_{i,j}^{n}}{| \bm{s}_{i,j}^{n} |} + \frac{\bm{l}_{i,j}^{n}}{| \bm{l}_{i,j}^{n} |} \nonumber \\
    \bm{\hat h}_{i,j}^{n} &= \frac{\bm{\hat l}_{i,j}^{n}}{| \bm{\hat l}_{i,j}^{n} |}
\end{align}
Therefore, this approach can be intuited as an extension of swarm optimization algorithms inspired by stochastic dynamics of variables belonging to sub-domains of $\mathbb R^n$ to the dynamics of a collection of spins that respectively belong to $\mathcal S^2$. Previous studies have applied a firefly optimization strategy to identifying noncollinear ground-states \cite{payneFireflyAlgorithmApplied2018}. However, one comparative benefit of the LLG approach, is that it precludes the need to explicitly define a distance metric between spin configurations, in addition to the fact that the LLG equation, and its variants, emerges in magnetic systems at multiple length-scales \cite{lakshmananFascinatingWorldLandau2011}. 

It is through this effective field that we may also introduce gradient information in order to improve convergence near the local minima. The constraining effective site magnetic field, $\bm{H}^{\text{eff}}_i$, can be described as the following \cite{maConstrainedDensityFunctional2015}
\begin{align}
    \bm{H}^{\text{eff}}_i &= 
    2 \lambda \left[{\bm{M}}_i -{\hat{\bm{M}}}_i^0\left({\hat{\bm{M}}}_i^0\cdot {\bm {M_i}}\right)\right]
\end{align}
where ${\bm{M}}_i$ are the integrated magnetic moments at site $i$, and ${\hat{\bm{M}}}_i^0$ are the unit vectors pointing in the individual site constraining directions \cite{maConstrainedDensityFunctional2015}.

\section{Results}

In order to test the effectiveness of the \spso{} noncollinear ground state optimizer, we chose a representative set of magnetic materials with exotically textured magnetic orderings from the Bilbao MAGNDATA database \cite{bilbao-server}. We first consider the convergence performance of \spso{} for a simple lattice Hamiltonian in Section~\ref{sec:ConvergeTest}.

\subsection{\spso{} convergence tests using model Hamiltonian}
\label{sec:ConvergeTest}

The improved convergence speed with the inclusion of gradient information is shown in \cref{fig:grad_compare_convergence}. We use a model classical Heisenberg Hamiltonian of the form
\begin{align}
    \mathcal H = - \sum_{i,j} J_{ij} \ \bm s_i \cdot \bm s_j, \quad \text{where } \bm s_i \in \mathcal S^2
    \label{eq:model_ham}
\end{align}
In order to test the convergence performance of the \spso{} algorithm, we employ the use of the Heisenberg model Hamiltonian in order to robustly quantify the statistics of the of the stochastic agent-based optimization scheme. Namely, we use this model to test how sensitive the \spso{} is to the input random configurations, number of agents, and other hyper-parameters. 

Perhaps most importantly, we quantify the increased local convergence by including gradient contributions to the GCPSO algorithm in \cref{fig:grad_compare_convergence}. In this plot, we compare averaged trajectories over 10 swarms initialized with initial spin configurations sampled uniformly over the surface of a unit sphere, with a comparison between 4 and 24 agents. In \cref{fig:grad_compare_convergence} we show the mean and standard deviation of these trajectories visualized using error-bars. By incorporating gradient information, the algorithm converges much more rapidly to the expected ferromagnetic or antiferromagnetic ground state compared to the conventional randomly sampled best-agent perturbations used in GCPSO. The latter does not even converge within limit of thirty iterations.

\subsection{Material test cases}

\subsubsection{\ce{MnPtGa}}
\label{sec:MnPtGa-Results}

We start with the \ce{MnPtGa} test case, which has been explored as a promising magnetocaloric material \cite{MnPtGa-exp}. Additionally, it has been identified as a promising skyrmion host \cite{MnPtGa-exp}. Using the \spso{} optimization scheme, with calls to the source-free noncollinear exchange correlation functional, we observe rapid convergence to the experimentally resolved magnetic ordering. In Ref.~\cite{mooreSourceFree}, this system was also explored for its ground-state convergence using the source-free functional. It was found that using the source-free functional, the moments converged robustly to the symmetric experimentally resolved canted ferromagnetic ordering. Therefore, in this study we explore the global convergence to the ground state. We use Crystal Toolkit \cite{hortonCrystalToolkitWeb2023} in \cref{fig:MnPtGa_struct_converge}, in which we superimpose the convergence path of one agent in the particle swarm optimization PSO swarm, where the color and magnitude are scaled according to the position along the convergence trajectory. 

\subsubsection{\ce{YMnO3} and \ce{FeF3}}
\label{sec:YMnO3-FeF3-Results}

Next, we test to hexagonal magnetic structures within the \spso{} framework. We test these structures because the hexagonal symmetry results in a complex magnetic ground-state with a relatively small unit cell. However, for these two materials \ce{YMnO3} and \ce{FeF3}, it was observed that the source-free functional, in combination with this \spso{} optimization scheme, resulted in a minimum energy configuration with ferromagnetic orderings of the spins in $ab$ planes, as shown in \cref{fig:YMnO3-compare-fm-ab}, compared to the experimental ground-state \cref{fig:YMnO3-compare-exp}. This magnetic ordering has supposedly been experimentally resolved, but only at temperatures above 75 Kelvin in \ce{ScMnO3} \cite{YMnO3-exp}. 

Under these assumptions, we observe very good convergence to the experimental ground state. A majority of the runs for \ce{YMnO3} converge to the correct state, as shown in \cref{fig:YMnO3-SpinPSOtoSF}, where the initial states on the left are the \spso{}+GGA converged structures, used as input to \Fsf{GGA}, which converges to the corresponding orderings on the right hand side. This improved local convergence using the source-free functional is not surprising, based on the analysis and results of past studies \cite{mooreSourceFree, krishnaCompleteDescriptionMagnetic2019, sharmaSourceFreeExchangeCorrelationMagnetic2018a}. 

\subsubsection{\ce{Mn3Pt} and \ce{Mn3Sn}}
\label{sec:Mn3Pt-Mn3Sn-Results}

\ce{Mn3Pt} and \ce{Mn3Sn} are also studied using \spso{}, followed by calls to \Fsf{GGA}. The resulting ground-states are reported in \cref{fig:Mn3Pt-SpinPSOtoSF} and \cref{fig:Mn3Sn-SpinPSOtoSF}. We see that compared to the experimentally $R\overline{3}m$ experimentally resolved structure \cite{Mn3Pt-exp}, the algorithm \spso{} finds a configuration with circulating spins to be more favorable. This is possibly due to the fact that the source-free functional imposes a strict dependence on the spin current, which is proportional to the curl of the magnetization $\nabla \times \bm m$ \cite{mooreSourceFree, sharmaSourceFreeExchangeCorrelationMagnetic2018a}. The role of additional corrections to the source-free functional are explored in Section~\ref{sec:SFchallenge}. 

A similar circulation of spins are observed in the case of \ce{Mn3Sn}, shown in \cref{fig:Mn3Sn-SpinPSOtoSF}. However, it may not be as fair a comparison to experiment, as the neutron diffraction in Ref.~\onlinecite{Mn3Sn-exp} was performed around 200 K, and it is ambiguous whether the $Cmc'm'$ or $Cm'cm'$ structures is the preferred ground-state \cite{Mn3Sn-exp}. 

\subsubsection{\ce{Fe3PO3O4} and \ce{Mn2SiO4}}
\label{sec:Fe3PO3O4-Mn2SiO4-Results}

Lastly, we explore \ce{Fe3PO3O4} as a another material test case for our model. While the magnetic configuration has exhibited a rich incommensurate spin spiral ground-state using neutron diffraction \cite{rossNanosizedHelicalMagnetic2015, tarneTuningAntiferromagneticHelical2017}, we will explore the corresponding commensurate ground-state for the purposes of this study. This is because, for reasons discussed in Ref.~\onlinecite{mooreSourceFree}, additional functionality is required to incorporate spin-spiral boundary conditions with the source-free constraint. However, in future studies, we plan to use this material as a test case for optimizing a $q$-spiral ordering over the configuration space of the $q$ wave vector, by incorporating this degree of freedom into the \spso{} framework. The commensurate ordering is of iron arranged within triads, which are antiferromagnetically ordered with respect to one another. We see that the \spso{}+GGA computed ground-state agrees with the experimentally characterized structure. We also optimized this structure using \spso{}+\Fsf{GGA}, and found that a slight canting of the triad resulted, which does not agree with the experimentally measured structure. 

\subsection{Discussion on challenges with source-free functional and \spso{}}
\label{sec:SFchallenge}

We found that larger $U$ and $J$ values resulted in stronger ferromagnetic exchange, particularly in the case of in-plane spins in \ce{YMnO3}. Therefore, for these structures, we ran the \spso{} algorithm with no Hubbard $U$ and Hund $J$ values applied to the transition metal $d$ states. Therefore, it would be worthwhile to assess the effect of these inter-site $V$ values, i.e. \FpUJ{DFT}+$V$, which necessarily affect magnetic exchange the resulting ground state \cite{mahajanImportanceIntersiteHubbard2021a, moslehBenchmarkDensityFunctional2023a, ederIntersiteCoulombInteraction1996a, mollerMagneticOrderPeriodic1993}. Whether or not these inter-site $V$ values promote FM or AFM exchange is dependent upon the underlying character of the correlated insulator, e.g. whether or not the TMO is a Mott versus charge transfer insulator \cite{ederIntersiteCoulombInteraction1996a, mollerMagneticOrderPeriodic1993}, and warrants further investigation. 

However, what's possibly more of note in the context of this study are possible issues with the global energy curvature using the source-free functional.  In reference \cite{mooreSourceFree}, some of the current authors explored the local convergence of the \ce{YMnO3} magnetic state starting from random perturbations around the experimentally resolved ground state, and found robust conversions to this experimentally accepted configuration over several sample trials.

Another potential shortcoming of the source-free constrained functional is the neglection of explicit coupling between the probability current and XC magnetic vector potential, which will arise in spin polarized systems \cite{mooreSourceFree, capelleSpinDensityFunctionalsCurrentDensity1997a}. The question is whether or not these probability (paramagnetic) currents will have a significant effect in the case of quenched orbital moments. The authors also explore the importance of the choice of $\overline{\bm B}_{xc}$, the integral of the source-free XC magnetic field. This will likely have a strong effect on the global energy landscape, based on the results from Ref.~\cite{mooreSourceFree}. 
% Lastly, the explanation could simply come down to the shortcomings of \FpUJ{GGA} in describing magnetism in strongly correlated systems \cite{}. Therefore, higher-order functionals may be worth pursuing in this context.

In addition, different couplings could be important to consider. It might be necessary in future studies for us to incorporate optimization over both structure and degrees of freedom using the \spso{} formalism. In reference \cite{mooreSourceFree}, a stronger dependence of the computed ground-state on structural geometry relaxations were observed for the \FsfUJ{GGA}. This more holistic approach may help to resolve some of the issues in the agreement or disagreement between experiment and computed magnetic orderings using \spso{} with \FsfUJ{DFT}.

\section{Conclusions}

In this study, we have developed and implemented a \spso{} workflow in the \texttt{atomate} software code framework. We've demonstrated its improved convergence capabilities, by including gradient information in the form of local fields from constrained DFT calculations. For several material systems of varying chemistries, we have obtained the correct magnetic ground-state based on neutron diffraction studies. The best results were found for \spso{} with calls to GGA during global convergence, followed by \Fsf{GGA} calculations to improve local convergence behavior.

The underlying reasons for the issues with \Fsf{GGA} in global convergence remain unclear. Furthermore, we have thoroughly tested the global convergence properties of the source-free functional \cite{mooreSourceFree}, many of which can likely be addressed by further improvements to the new XC functional, which are proposed in Ref.~\onlinecite{mooreSourceFree}.

This work can be further extended to optimize over spin-spiral degrees of freedom, which would include a $q$-spiral vector and a spin-quantization axis. These boundary conditions are necessary to describe magnetic systems with an incommensurate magnetic ground-state, and to reduce computational cost more generally. These extensions will require further care, such as the accurate and computationally efficient manner in which to compute the corresponding gradients associated with the $q$-vector. In principle, this could be achieved by leveraging the Fourier scaling property.

% \vspace{2ex}

\section{Acknowledgements}

G.M. would like to thank Caitlin McCandler and Evan Spotte-Smith for their support; the three of them learned about and benchmarked particle swarm optimization algorithms together during the final project for a parallel computing course offered by the UC Berkeley Electrical Engineering and Computer Sciences department.
% course CS267 
% executed on distributed memory computing architectures 
G.M. acknowledges support from the Department of Energy Computational Science Graduate Fellowship (DOE CSGF) under grant DE-SC0020347. Computations in this paper were performed using resources of the National Energy Research Scientific Computing Center (NERSC), a U.S. Department of Energy Office of Science User Facility operated under contract no. DE-AC02-05CH11231. Expertise in high-throughput calculations, data and software infrastructure was supported by the U.S. Department of Energy, Office of Science, Office of Basic Energy Sciences, Materials Sciences and Engineering Division under Contract DE-AC02-05CH11231: Materials Project program KC23MP.

\begin{widetext}

\begin{figure}[h]
    \centering
    \includegraphics[width=0.85\linewidth]{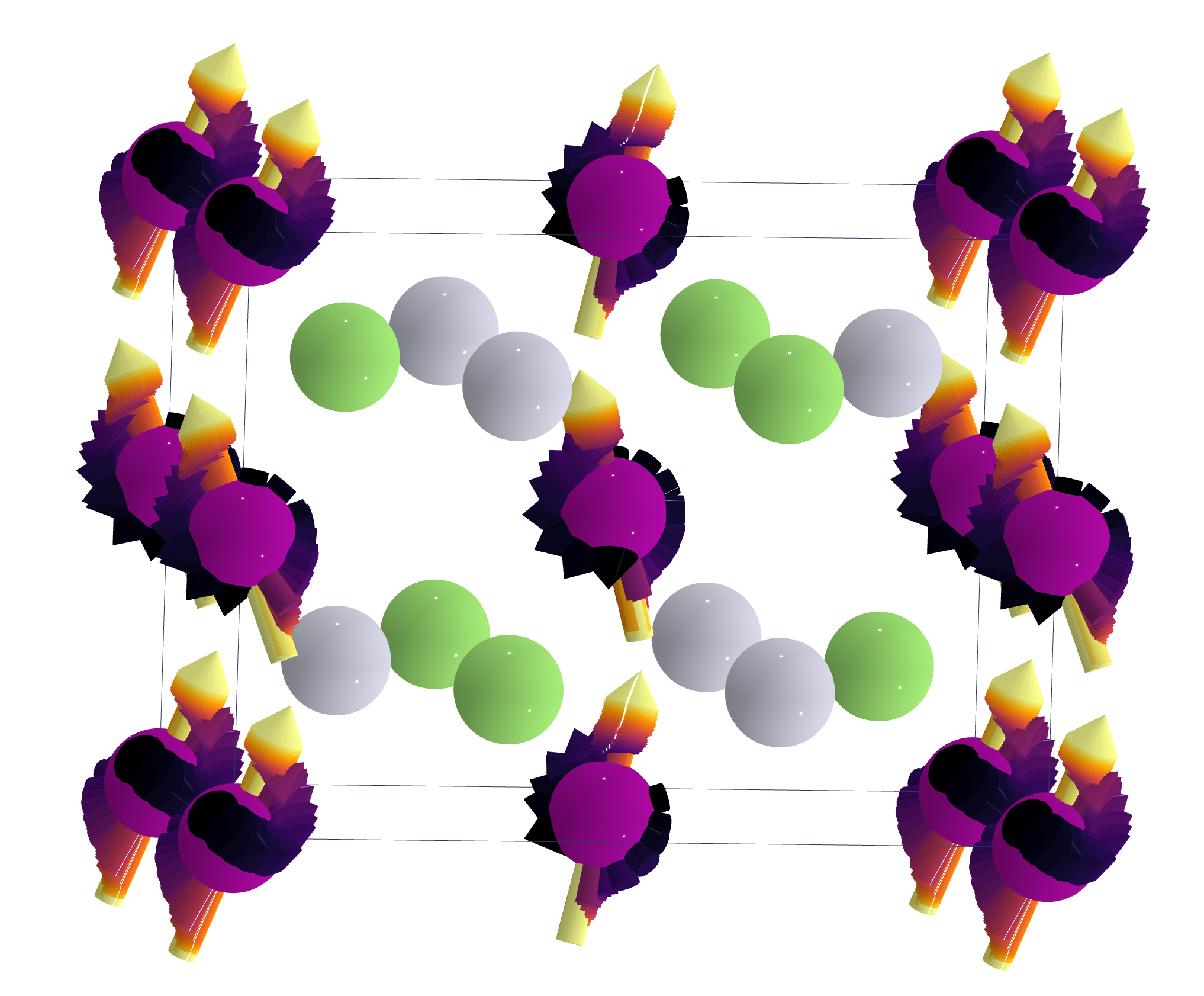} 
    \caption{The convergence path for one of four agent's trajectory randomly initialized within a single \spso{} optimization, with calls to \FsfUJ{PBE}. The algorithm converges within fifteen iterations to the experimentally resolved magnetic ground-state \cite{MnPtGa-exp}. The position along the trajectory is indicated by the length and color of the arrows within the ``inferno" color scheme. }
    \label{fig:MnPtGa_struct_converge}
\end{figure}

\begin{figure}[h]
    \centering
    \includegraphics[width=0.85\linewidth]{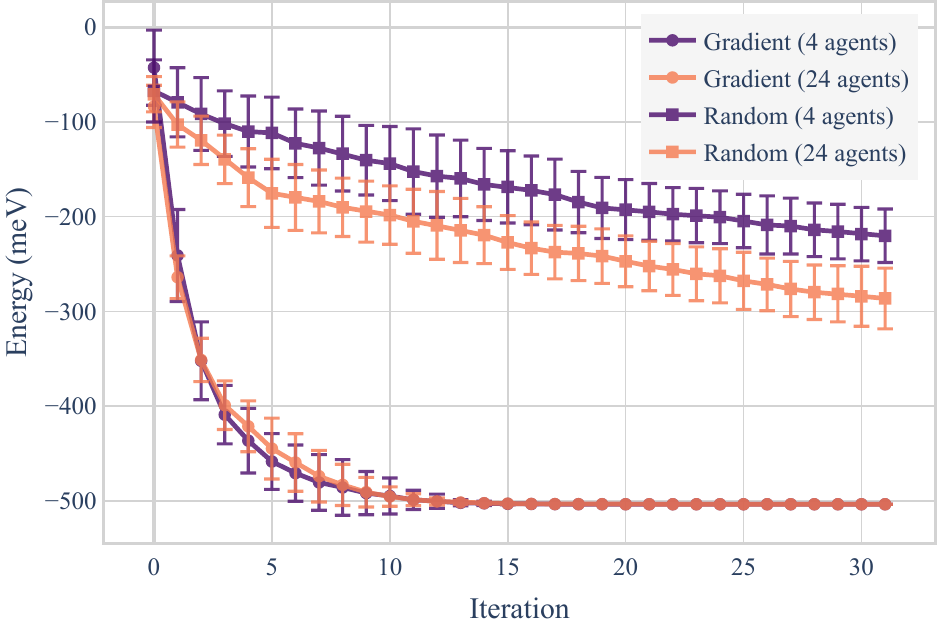} 
    \caption{The convergence of \spso{} with and without gradient information, as stated in \ref{}. For the sake of computational efficiency, energies and local effective fields are evaluated using \cref{eq:model_ham} for a Heusler ferrimagnet. Additionally, to explore the sensitivity of each setting to hyperparameters, we compare convergence with 4 versus 24 agents, and averaged over 10 randomly initialized swarms.}
    \label{fig:grad_compare_convergence}
\end{figure}

\begin{figure}[h]
    \centering
    \subfloat[\centering Convergence with 6 agents]{{
    %\label{fig:}
    \includegraphics[width=0.90\linewidth]{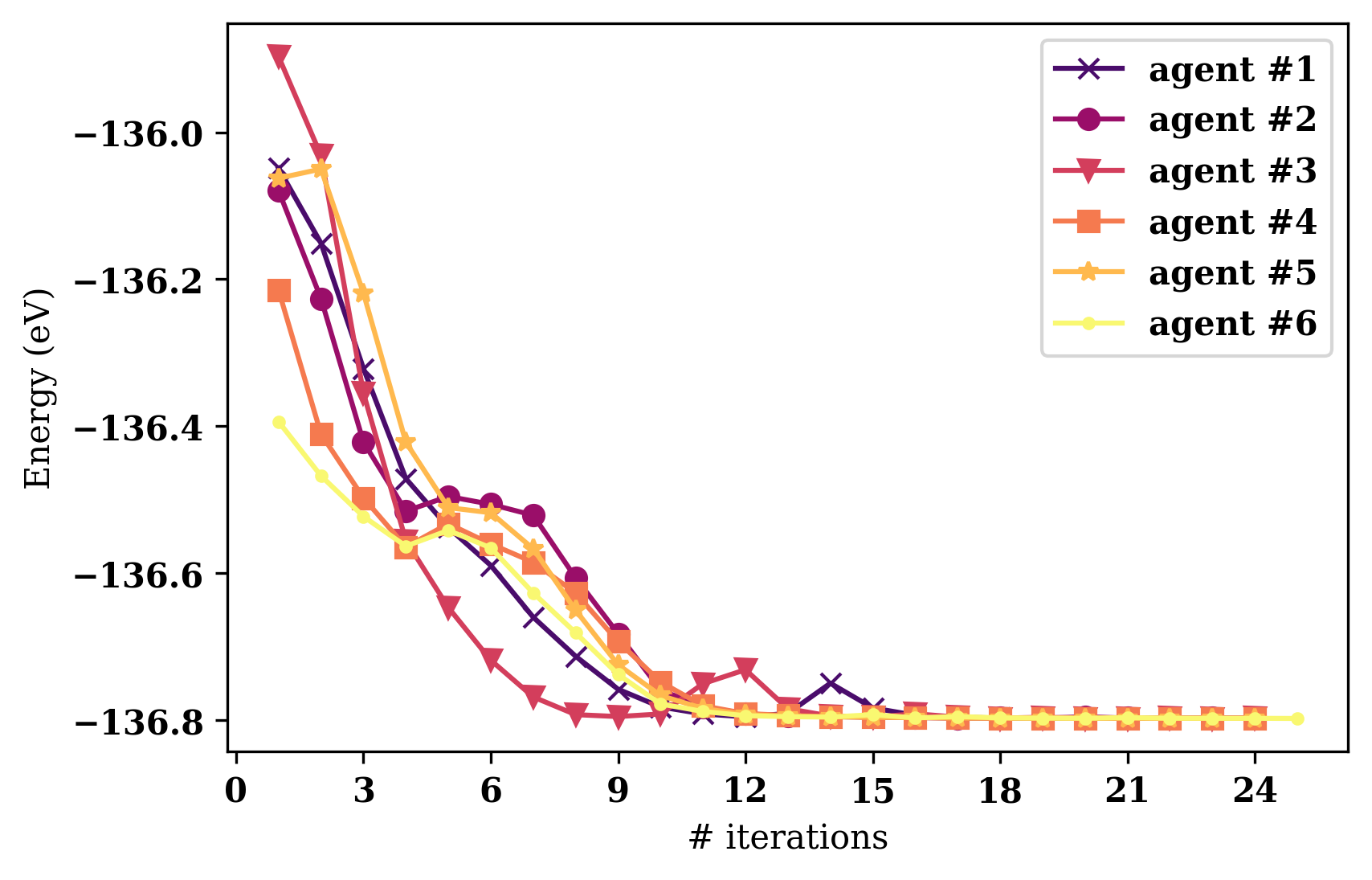}}} \\
    \subfloat[\centering Convergence with 12 agents]{{
    %\label{fig:}
    \includegraphics[width=0.90\linewidth]{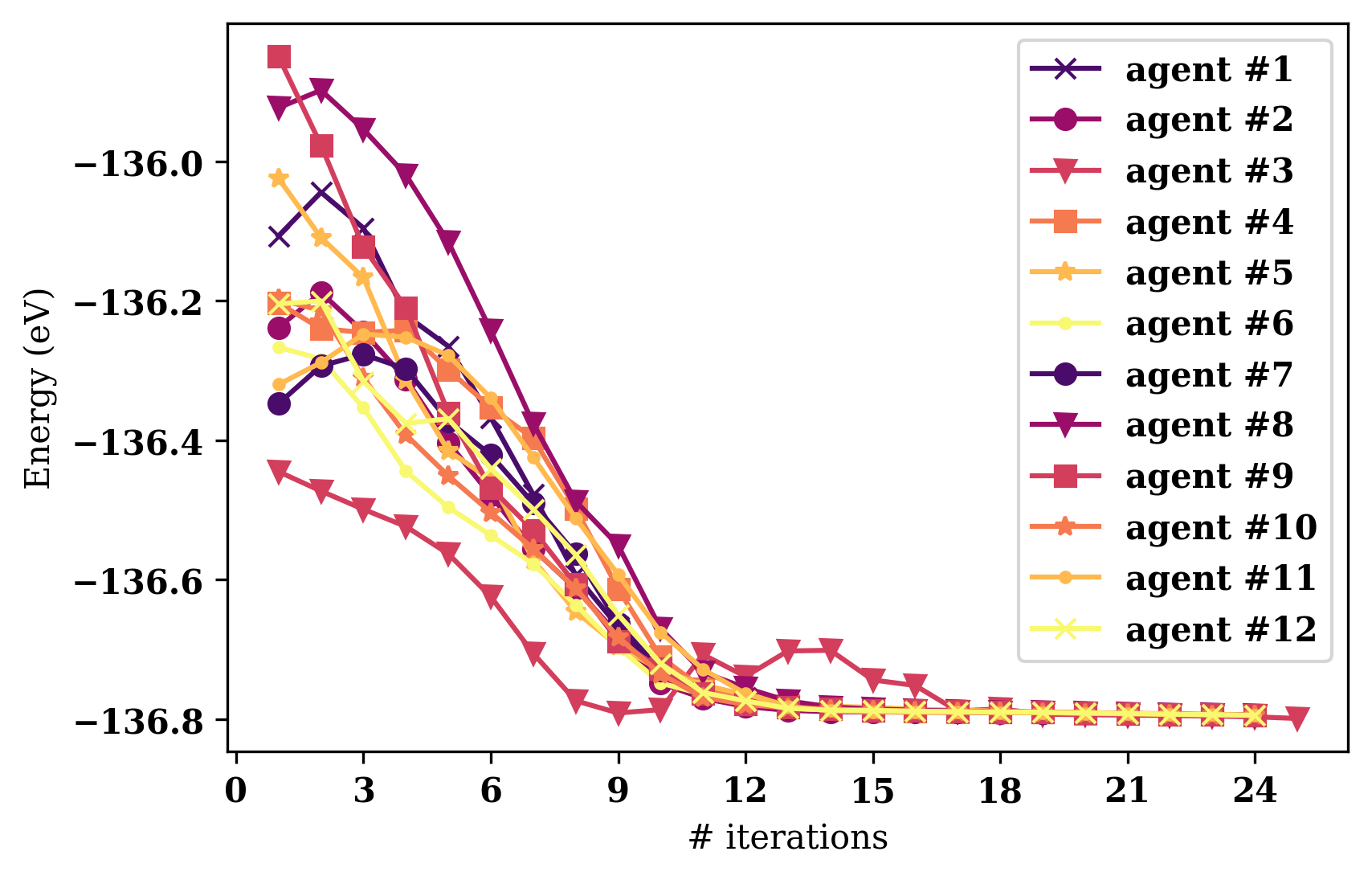}}}
    \caption{ The robust convergence for two randomly initialized \spso{} trajectories for the \ce{FeF3} magnetic ground-state. This figure provides a convergence comparison for swarm sizes of (a) six agents versus (b) twelve. \cref{fig:grad_compare_convergence} provides a statistically averaged comparison over agent sizes, which is only computationally practical using the model Hamiltonian, \cref{eq:model_ham}. }
    \label{fig:FeF3convergence}
\end{figure}

\begin{figure}[h]
    \centering
    \includegraphics[width=0.98\linewidth]{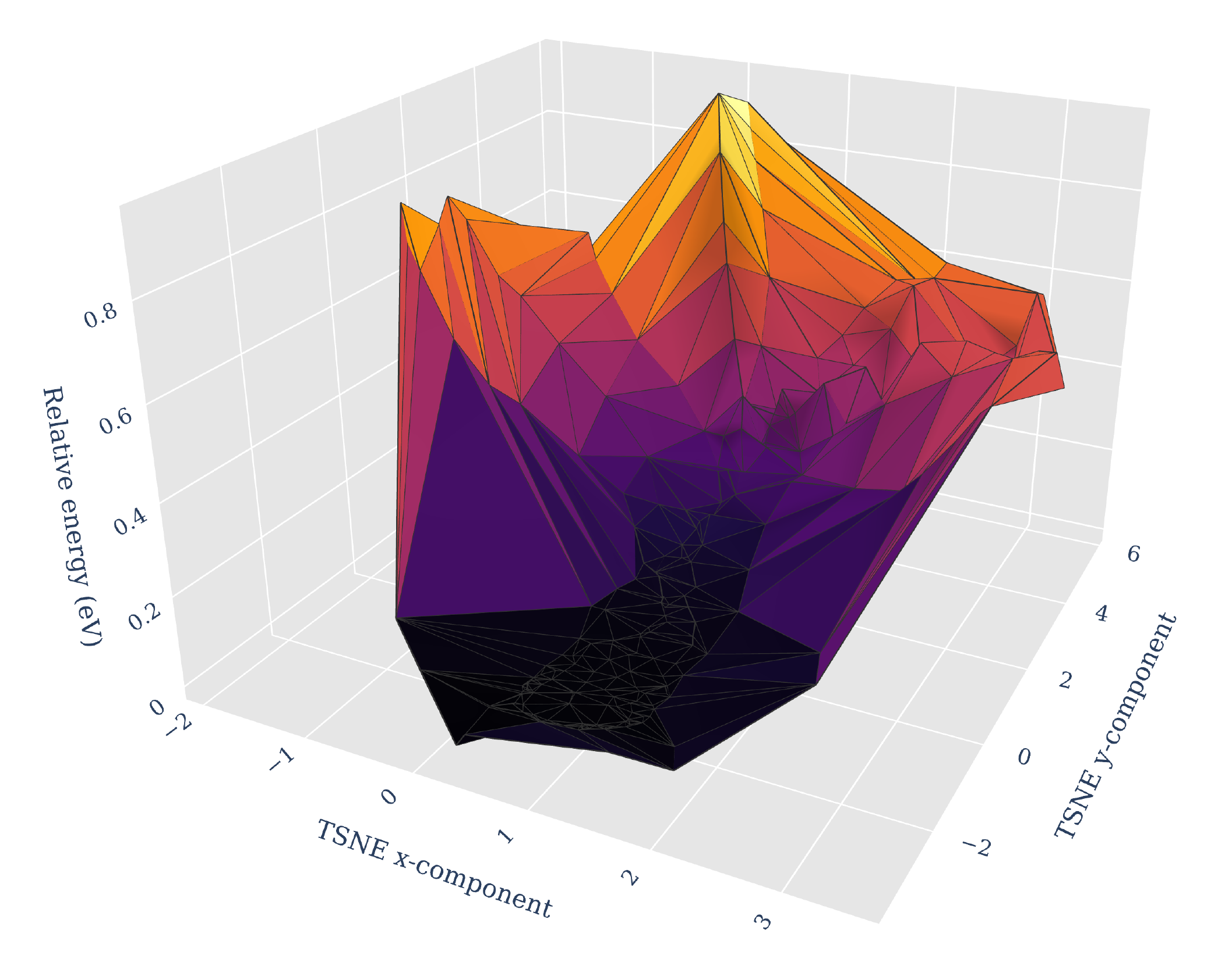} 
    \caption{ The energy landscape for \ce{FeF3} spin configurations, visualized using a $t$-distributed stochastic neighbor embedding (acronymed TSNE or t-SNE) \cite{hintonStochasticNeighborEmbedding2002} implemented in \texttt{sklearn}. The energy values and positions (spin configurations) are amalgamated from each agent's trajectory within the swarm. }
    \label{fig:Esurf-FeF3}
\end{figure}

\begin{figure}[h]
    \centering
    %%
    %\begin{mdframed}[roundcorner=10pt, linewidth=1.5pt]
    \centering
    {\large Trial I} \\
    \vspace{2ex}
    \begin{minipage}{0.49\linewidth}
    \begin{mdframed}[roundcorner=10pt, linewidth=1.5pt]
        \centering
        \textit {Converged structure:} \\ PBE \\
        \subfloat[\centering PBE, trial I \newline \textit{viewed along [001]}]{{
        %\label{fig:}
        \includegraphics[width=0.54\linewidth]{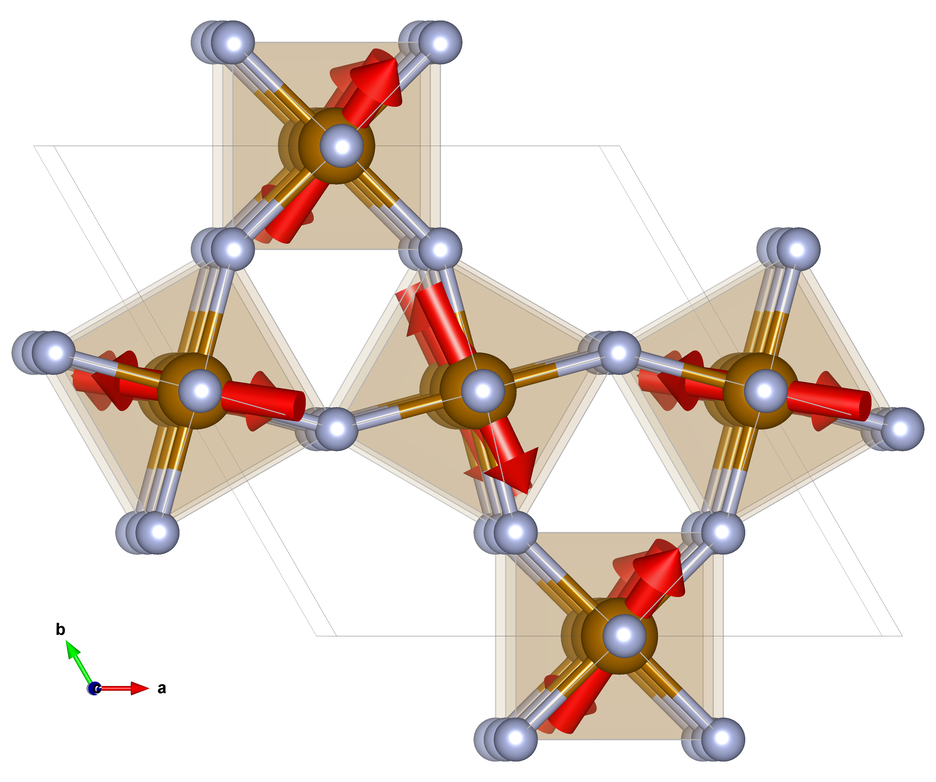}}}
        \subfloat[\centering PBE, trial I \newline \textit{viewed along [100]}]{{
        %\label{fig:}
        \includegraphics[width=0.44\linewidth]{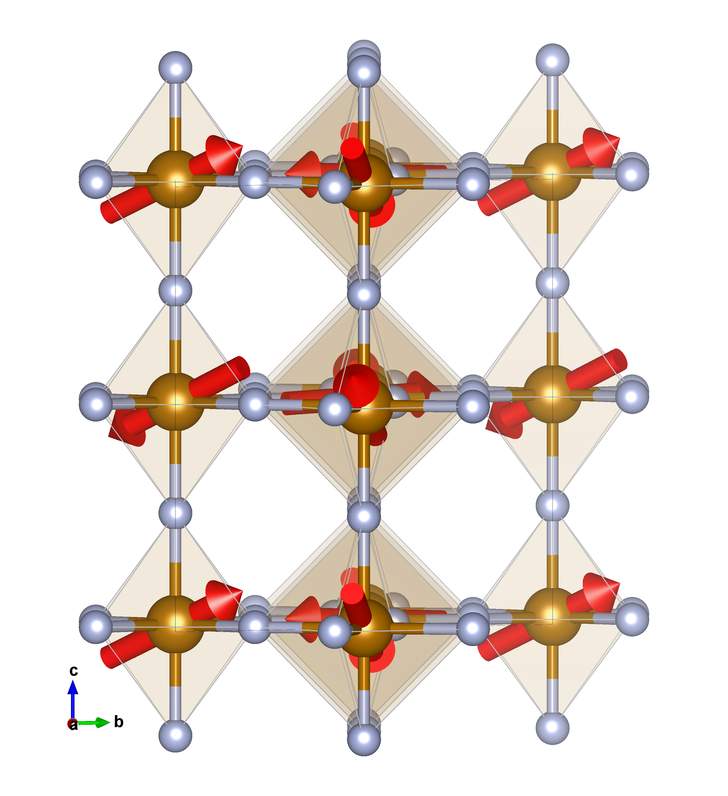}}}
    \end{mdframed}
    \end{minipage}
    \begin{minipage}{0.49\linewidth}
    \begin{mdframed}[roundcorner=10pt, linewidth=1.5pt]
        \centering
        \textit {Converged structure:} \\ Source-free PBE \\
        \subfloat[\centering PBE\textsubscript{SF}, trial I \newline \textit{viewed along [001]}]{{
        % \label{fig:}
        \includegraphics[width=0.54\linewidth]{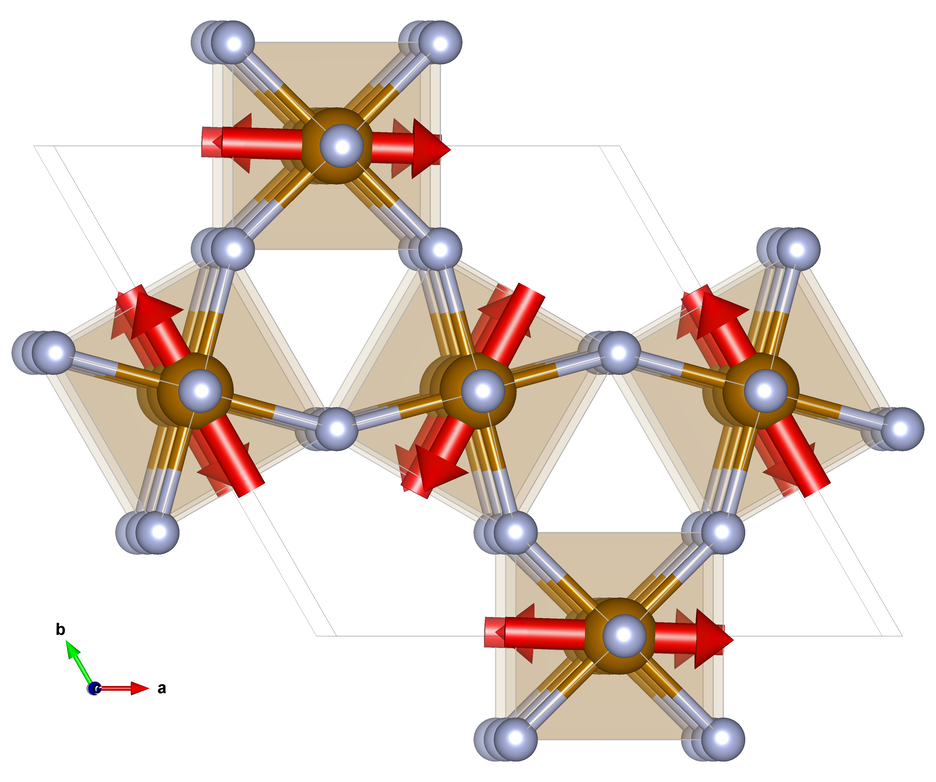}}}
        \subfloat[\centering PBE\textsubscript{SF}, trial I \newline \textit{viewed along [100]}]{{
        % \label{fig:}
        \includegraphics[width=0.44\linewidth]{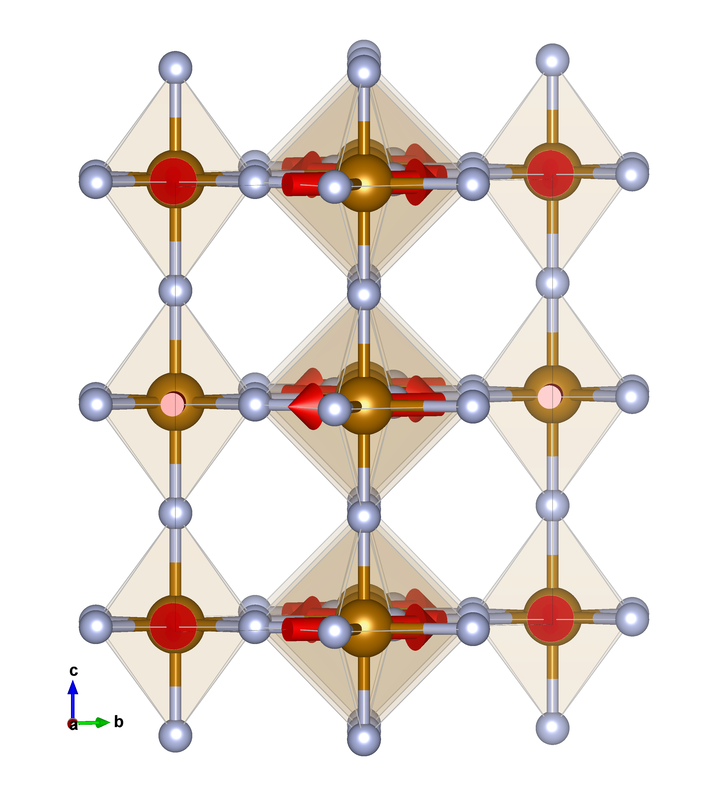}}}
    \end{mdframed}
    \end{minipage}
    %\end{mdframed}
    %%
    %\begin{mdframed}[roundcorner=10pt, linewidth=1.5pt]
    \centering
    \vspace{2ex} \\
    {\large Trial II} \\
    \vspace{2ex}
    \begin{minipage}{0.49\linewidth}
    \begin{mdframed}[roundcorner=10pt, linewidth=1.5pt]
        \centering
        \textit {Converged structure:} \\ PBE \\
        \subfloat[\centering PBE, trial II \newline \textit{viewed along [001]}]{{
        %\label{fig:}
        \includegraphics[width=0.54\linewidth]{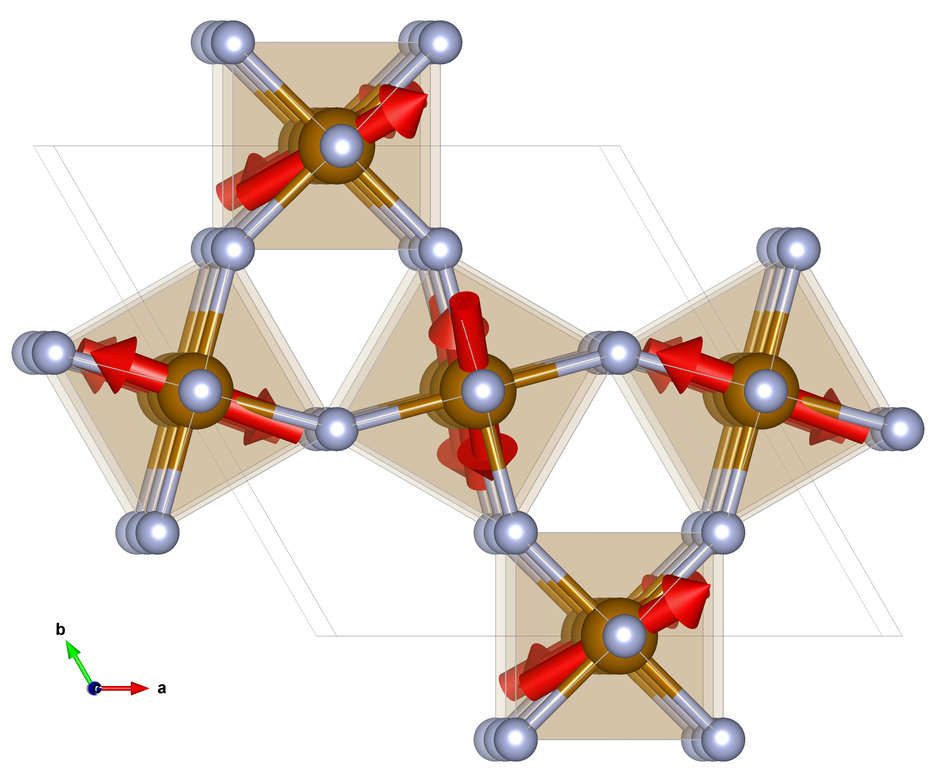}}}
        \subfloat[\centering PBE, trial II \newline \textit{viewed along [100]}]{{
        %\label{fig:}
        \includegraphics[width=0.44\linewidth]{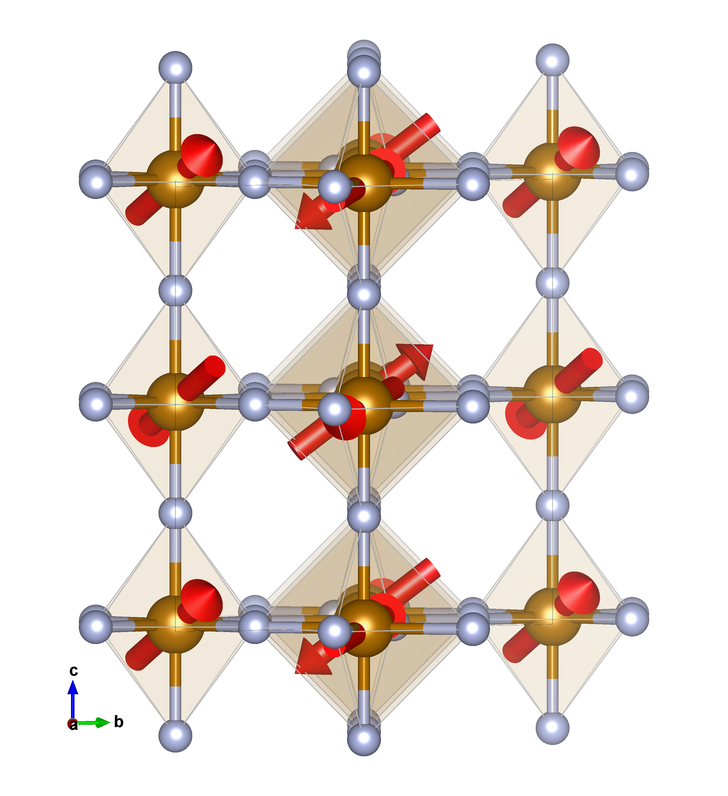}}}
    \end{mdframed}
    \end{minipage}
    \begin{minipage}{0.49\linewidth}
    \begin{mdframed}[roundcorner=10pt, linewidth=1.5pt]
        \centering
        \textit {Converged structure:} \\ Source-free PBE \\
        \subfloat[\centering PBE\textsubscript{SF}, trial II \newline \textit{viewed along [001]}]{{
        % \label{fig:}
        \includegraphics[width=0.54\linewidth]{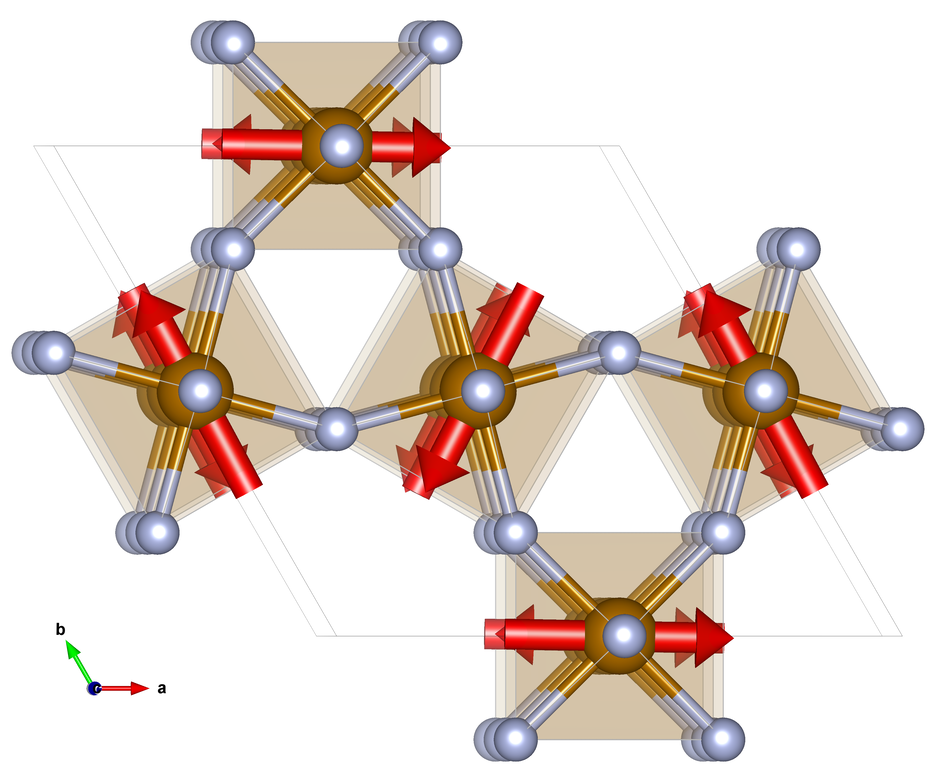}}}
        \subfloat[\centering PBE\textsubscript{SF}, trial II \newline \textit{viewed along [100]}]{{
        % \label{fig:}
        \includegraphics[width=0.44\linewidth]{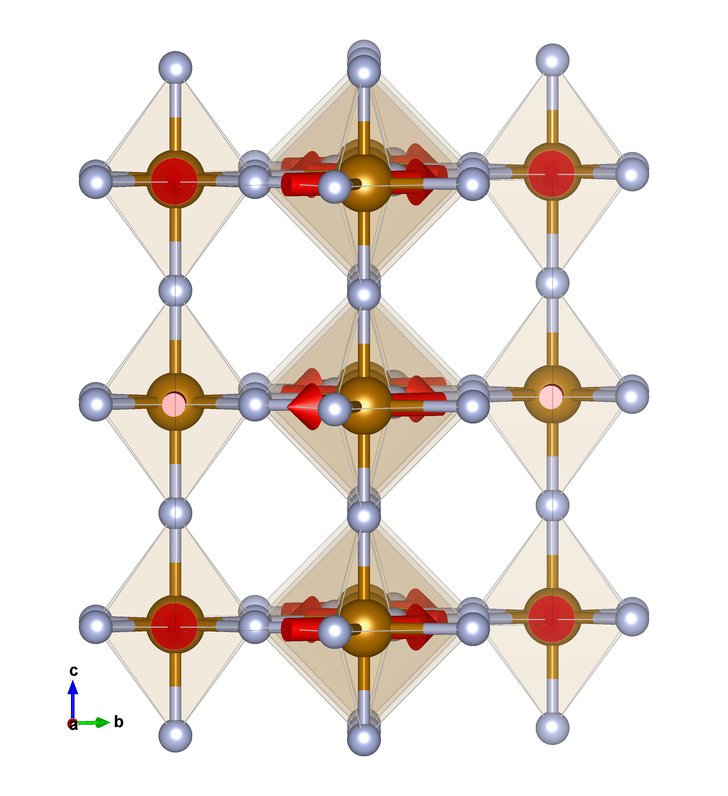}}}
    \end{mdframed}
    \end{minipage}
    %\end{mdframed}
    %%
    \caption{ \SpinPSOtoSFcaption{FeF3} }
    \label{fig:FeF3-SpinPSOtoSF}
\end{figure}

\begin{figure}[h]
    \centering
    %%
    %\begin{mdframed}[roundcorner=10pt, linewidth=1.5pt]
    \centering
    {\large Trial I} \\
    \vspace{2ex}
    \begin{minipage}{0.49\linewidth}
    \begin{mdframed}[roundcorner=10pt, linewidth=1.5pt]
        \centering
        \textit {Converged structure:} \\ PBE \\
        \subfloat[\centering PBE, trial I \newline \textit{viewed along [001]}]{{
        %\label{fig:}
        \includegraphics[width=0.64\linewidth]{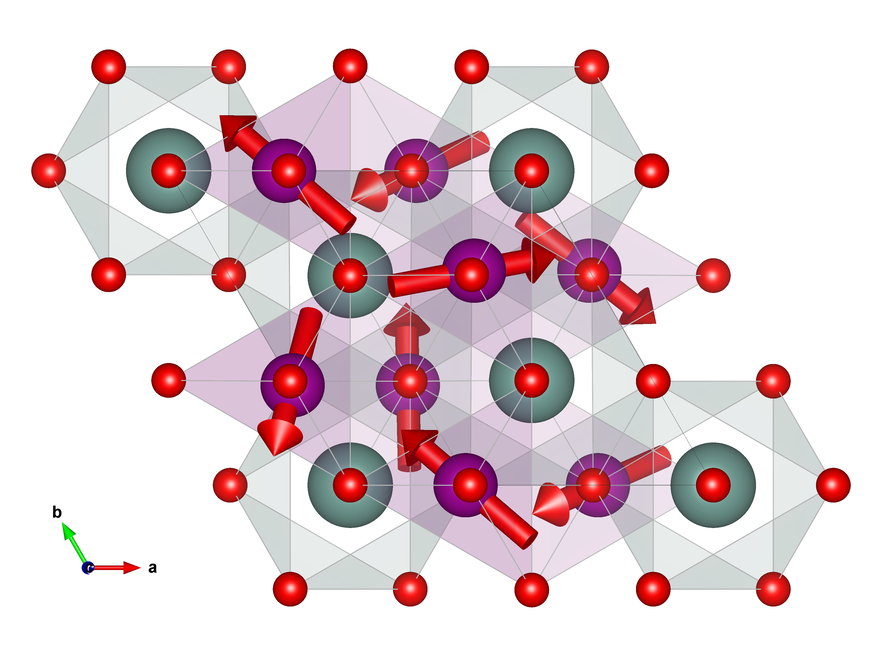}}}
        \subfloat[\centering PBE, trial I \newline \textit{viewed along [100]}]{{
        %\label{fig:}
        \includegraphics[width=0.34\linewidth]{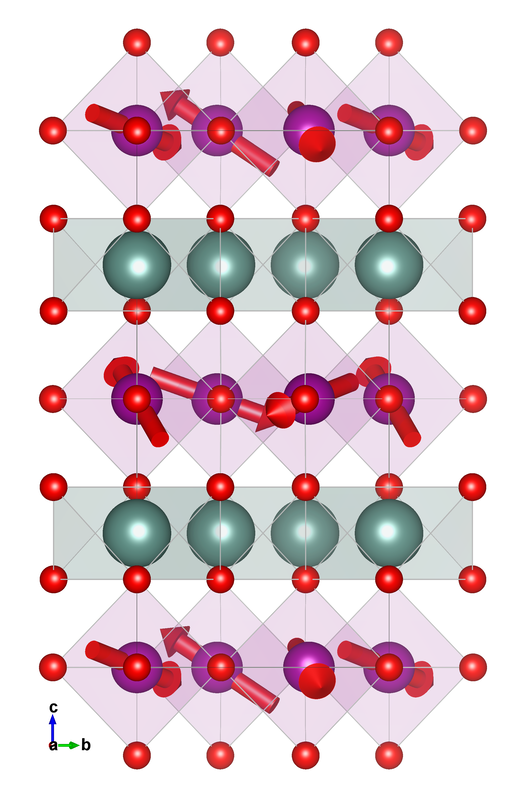}}}
    \end{mdframed}
    \end{minipage}
    \begin{minipage}{0.49\linewidth}
    \begin{mdframed}[roundcorner=10pt, linewidth=1.5pt]
        \centering
        \textit {Converged structure:} \\ Source-free PBE \\
        \subfloat[\centering PBE\textsubscript{SF}, trial I \newline \textit{viewed along [001]}]{{
        % \label{fig:}
        \includegraphics[width=0.64\linewidth]{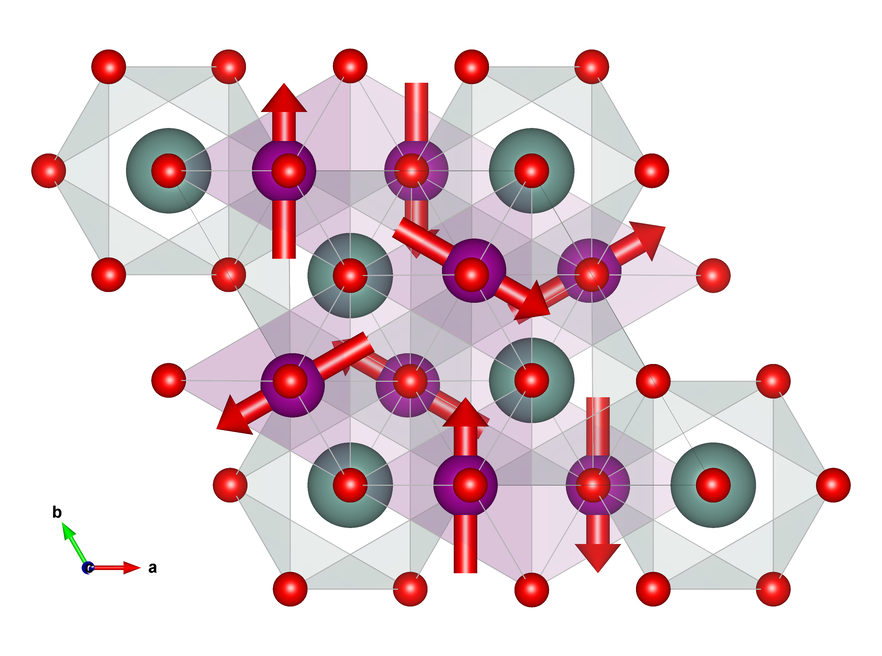}}}
        \subfloat[\centering PBE\textsubscript{SF}, trial I \newline \textit{viewed along [100]}]{{
        % \label{fig:}
        \includegraphics[width=0.34\linewidth]{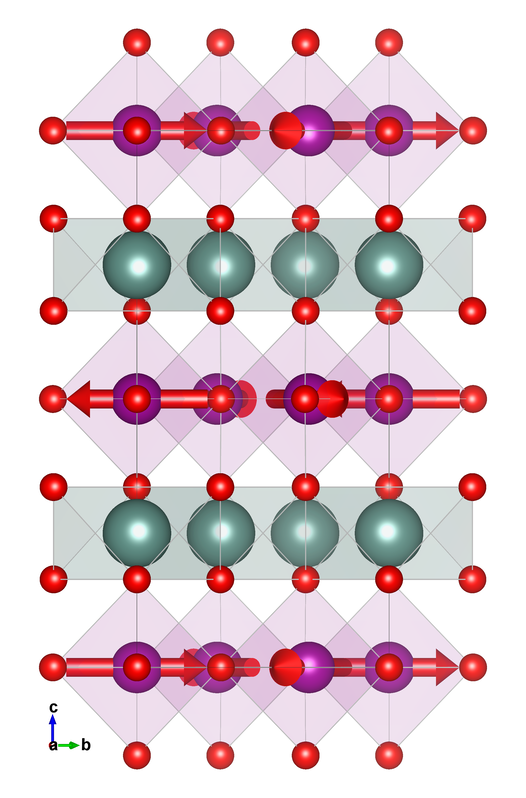}}}
    \end{mdframed}
    \end{minipage}
    %\end{mdframed}
    %%
    %\begin{mdframed}[roundcorner=10pt, linewidth=1.5pt]
    \centering
    \vspace{2ex} \\
    {\large Trial II} \\
    \vspace{2ex}
    \begin{minipage}{0.49\linewidth}
    \begin{mdframed}[roundcorner=10pt, linewidth=1.5pt]
        \centering
        \textit {Converged structure:} \\ PBE \\
        \subfloat[\centering PBE, trial II \newline \textit{viewed along [001]}]{{
        %\label{fig:}
        \includegraphics[width=0.64\linewidth]{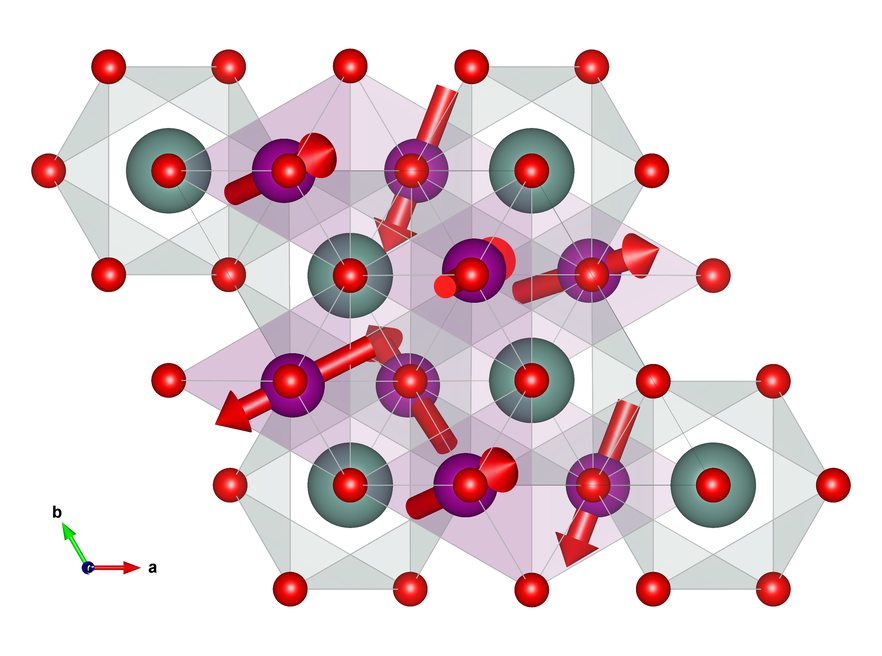}}}
        \subfloat[\centering PBE, trial II \newline \textit{viewed along [100]}]{{
        %\label{fig:}
        \includegraphics[width=0.34\linewidth]{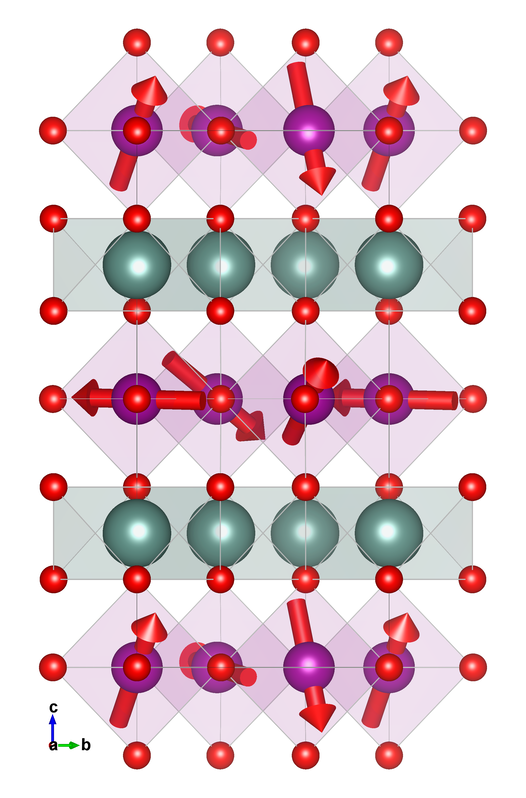}}}
    \end{mdframed}
    \end{minipage}
    \begin{minipage}{0.49\linewidth}
    \begin{mdframed}[roundcorner=10pt, linewidth=1.5pt]
        \centering
        \textit {Converged structure:} \\ Source-free PBE \\
        \subfloat[\centering PBE\textsubscript{SF}, trial II \newline \textit{viewed along [001]}]{{
        % \label{fig:}
        \includegraphics[width=0.64\linewidth]{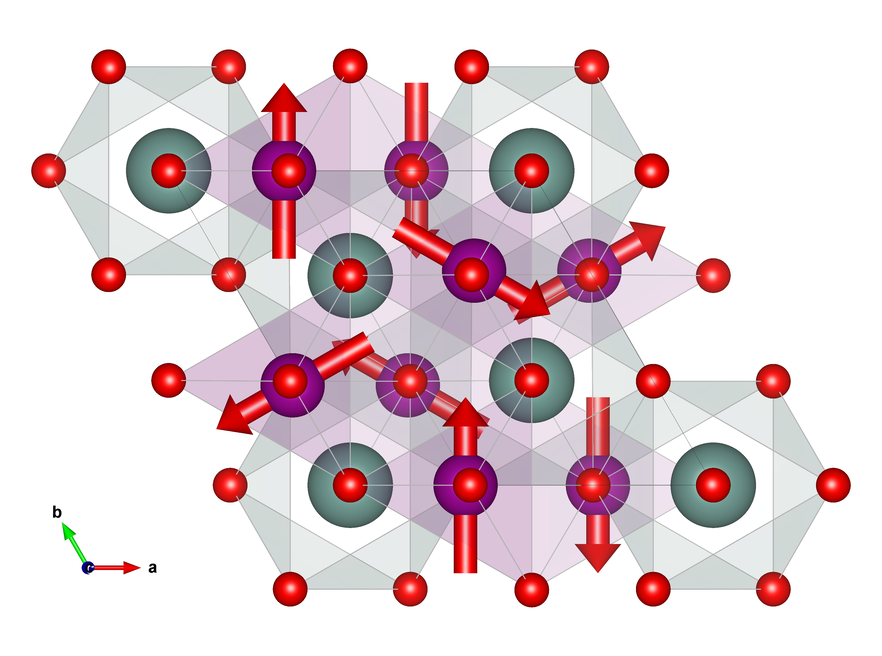}}}
        \subfloat[\centering PBE\textsubscript{SF}, trial II \newline \textit{viewed along [100]}]{{
        % \label{fig:}
        \includegraphics[width=0.34\linewidth]{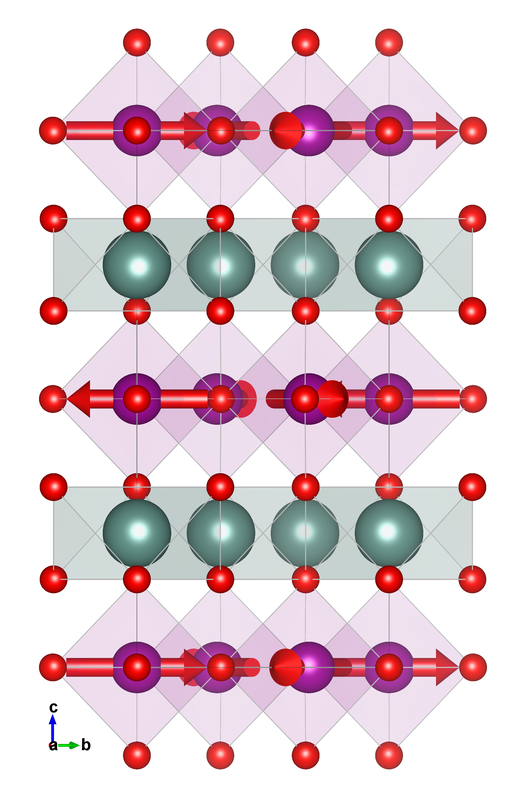}}}
    \end{mdframed}
    \end{minipage}
    %\end{mdframed}
    %%
    %\begin{mdframed}[roundcorner=10pt, linewidth=1.5pt]
    \centering
    \vspace{2ex} \\
    {\large Trial III} \\
    \vspace{2ex}
    \begin{minipage}{0.49\linewidth}
    \begin{mdframed}[roundcorner=10pt, linewidth=1.5pt]
        \centering
        \textit {Converged structure:} \\ PBE \\
        \subfloat[\centering PBE, trial III \newline \textit{viewed along [001]}]{{
        %\label{fig:}
        \includegraphics[width=0.64\linewidth]{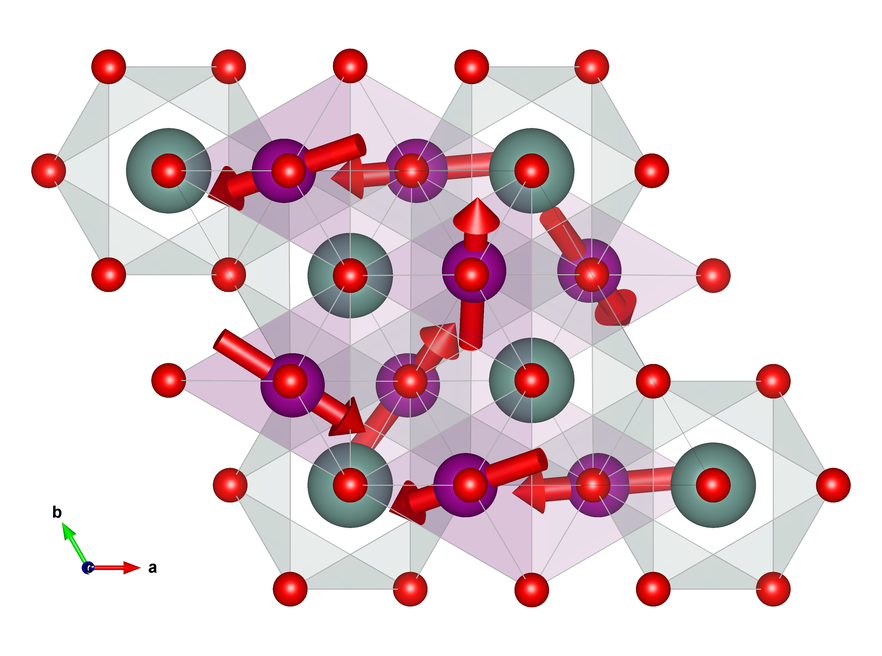}}}
        \subfloat[\centering PBE, trial III \newline \textit{viewed along [100]}]{{
        %\label{fig:}
        \includegraphics[width=0.34\linewidth]{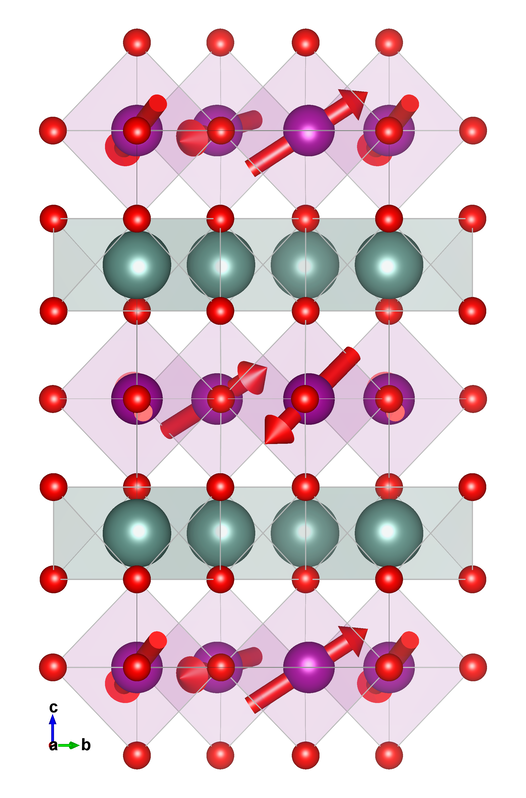}}}
    \end{mdframed}
    \end{minipage}
    \begin{minipage}{0.49\linewidth}
    \begin{mdframed}[roundcorner=10pt, linewidth=1.5pt]
        \centering
        \textit {Converged structure:} \\ Source-free PBE \\
        \subfloat[\centering PBE\textsubscript{SF}, trial III \newline \textit{viewed along [001]}]{{
        % \label{fig:}
        \includegraphics[width=0.64\linewidth]{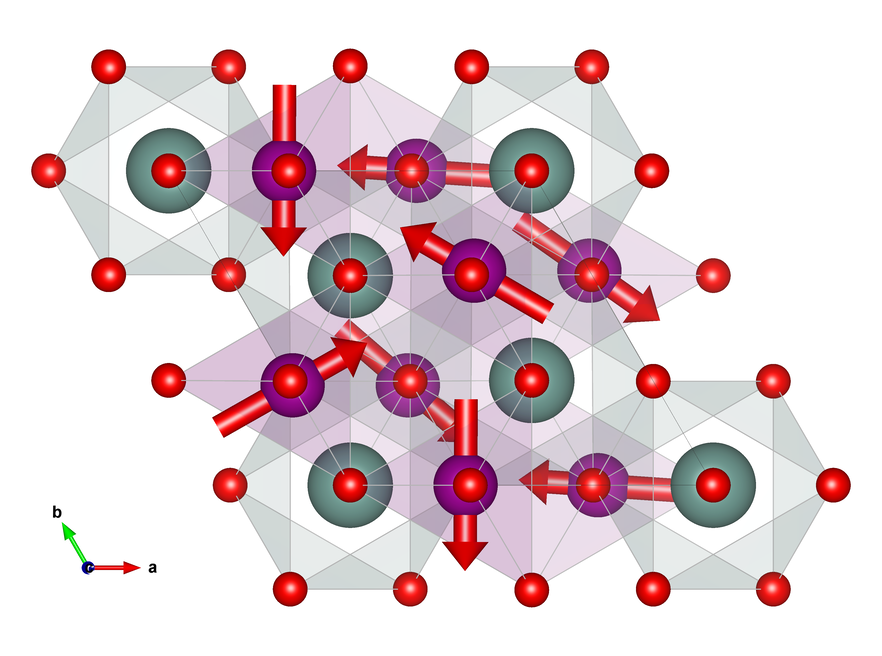}}}
        \subfloat[\centering PBE\textsubscript{SF}, trial III \newline \textit{viewed along [100]}]{{
        % \label{fig:}
        \includegraphics[width=0.34\linewidth]{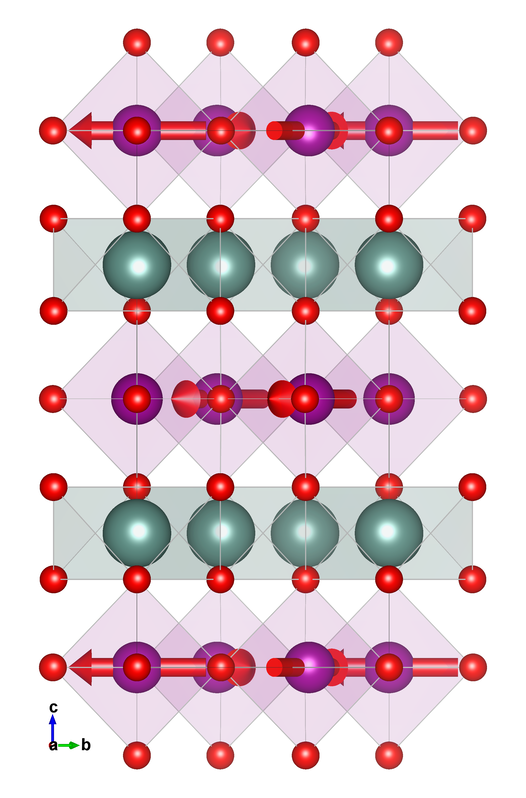}}}
    \end{mdframed}
    \end{minipage}
    %\end{mdframed}
    %%
    \caption{ \SpinPSOtoSFcaption{YMnO3} }
    \label{fig:YMnO3-SpinPSOtoSF}
\end{figure}

\begin{figure}[h]
    \centering
    \subfloat[\centering \FsfUJ{PBE}]{{
    \label{fig:YMnO3-compare-fm-ab}
    \includegraphics[width=0.48\linewidth]{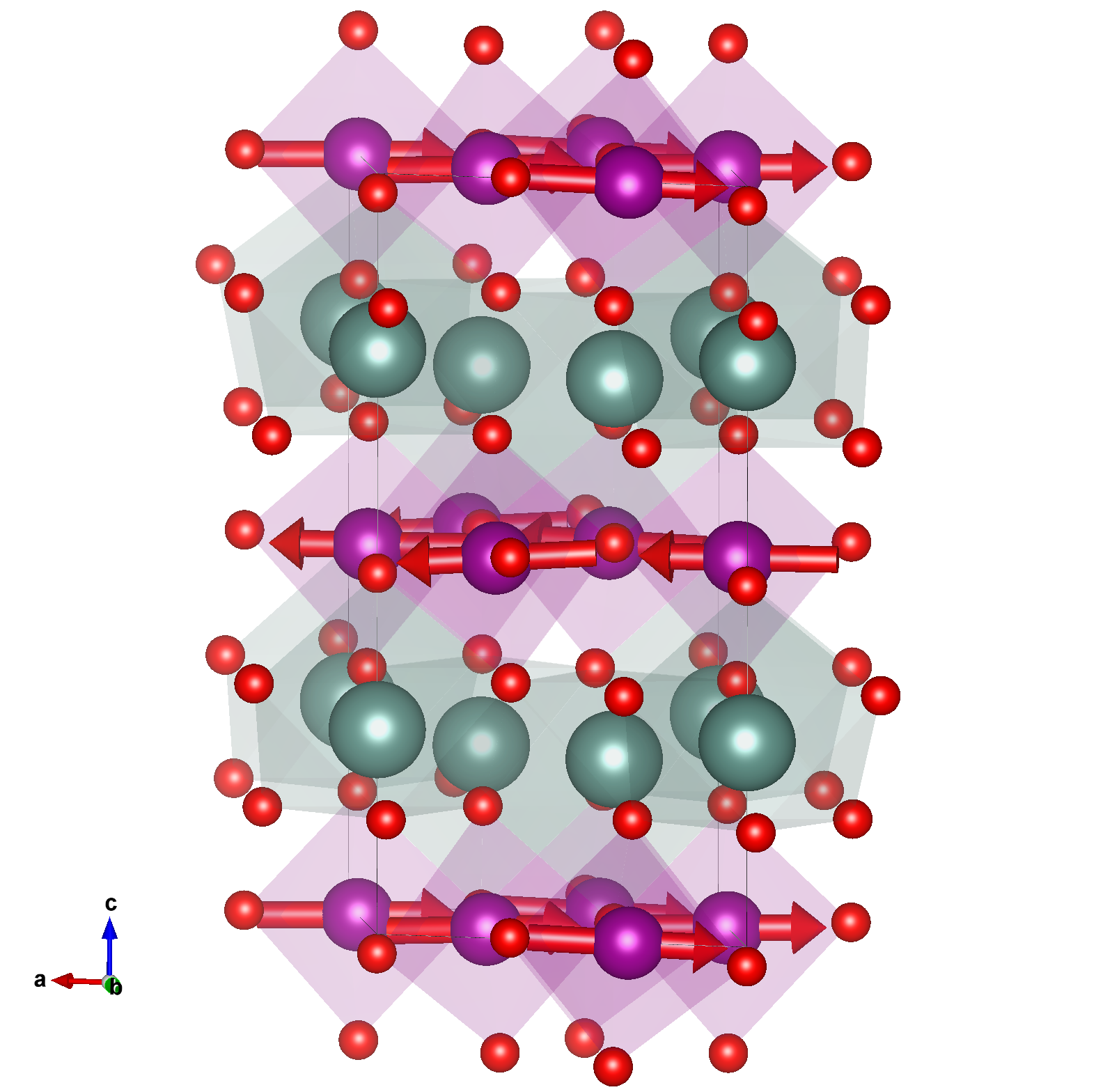}
    }}
    \subfloat[\centering \Fsf{PBE}]{{
    \label{fig:YMnO3-compare-exp}
    \includegraphics[width=0.48\linewidth]{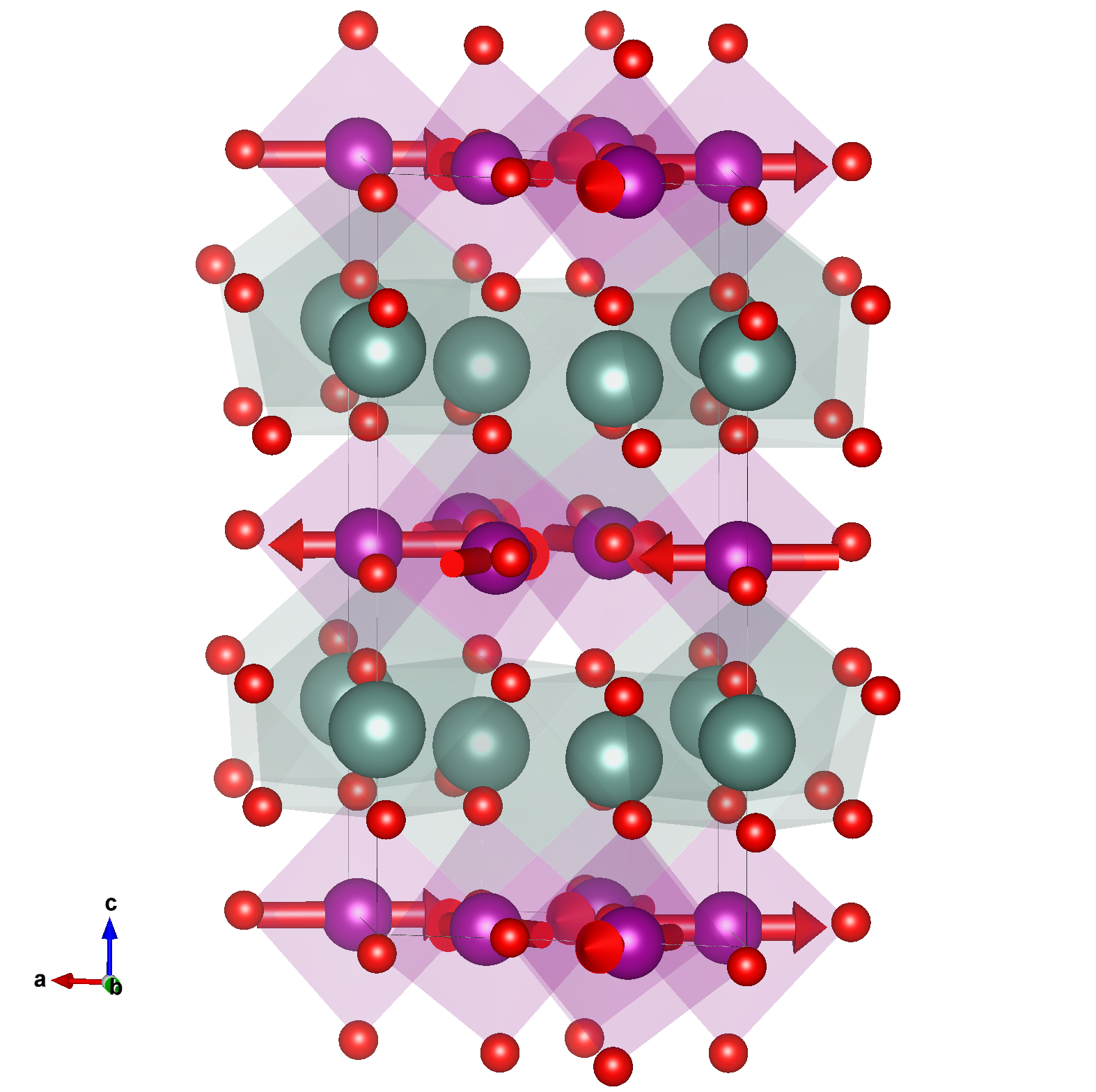}
    }}
    \caption{ (a) The \ce{YMnO3} ground state calculated using \spso{} with \FpUJ{PBE}, followed by \FpUJ{PBE} average Mn-$d$ Hubbard $U$ and Hund $J$ parameters obtained from Ref.~\onlinecite{mooreHighthroughputDeterminationHubbard2022a}. The calculated ground-state in this case contains ferromagnetic $ab$ planes, antiferromagnetically coupled. This structure differs from (b) the experimental structure \cite{YMnO3-exp}, as well as the ordering obtained from the procedure without $U$/$J$ parameters, using parameters consistent with \cref{fig:YMnO3-SpinPSOtoSF}. }
    \label{fig:YMnO3-compare}
\end{figure}

\begin{figure}[h]
    \centering
    %\begin{mdframed}[roundcorner=10pt, linewidth=1.5pt]
    \centering
    {\large Trial I} \\
    \vspace{2ex}
    \begin{minipage}{0.39\linewidth}
    \begin{mdframed}[roundcorner=10pt, linewidth=1.5pt]
        \centering
        \textit{Converged structure:} \\
        PBE \\
        \subfloat[\centering PBE, trial I]{{
        % \label{fig:}
        \includegraphics[width=0.80\linewidth]{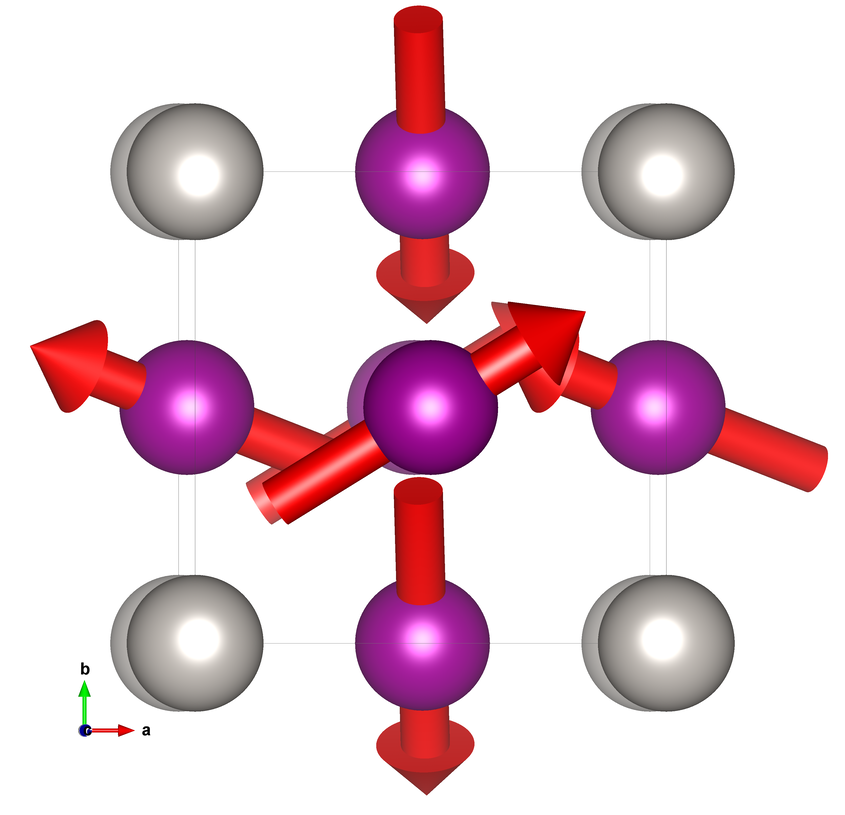}}}
    \end{mdframed}
    \end{minipage}
    \begin{minipage}{0.39\linewidth}
    \begin{mdframed}[roundcorner=10pt, linewidth=1.5pt]
        \centering
        \textit{Converged structure:} \\
        Source-free PBE \\
        \subfloat[\centering \Fsf{PBE}, trial I]{{
        %\label{fig:}
        \includegraphics[width=0.80\linewidth]{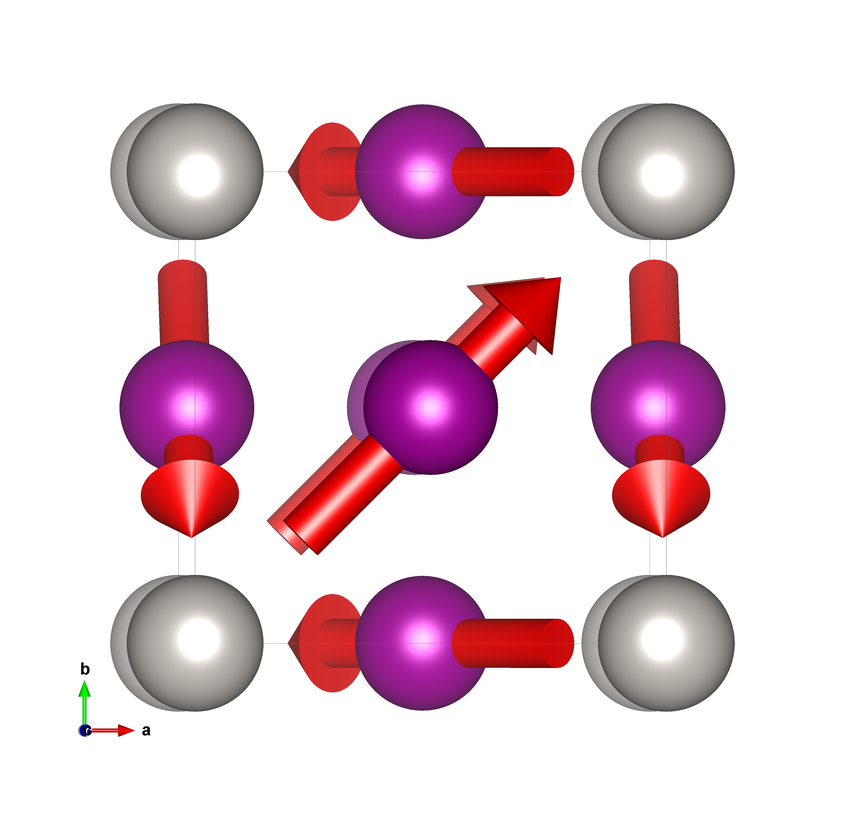}}}
    \end{mdframed}
    \end{minipage}
    %\end{mdframed}
    %%
    %\begin{mdframed}[roundcorner=10pt, linewidth=1.5pt]
    \centering
    \vspace{2ex} \\
    {\large Trial II} \\
    \vspace{2ex}
    \begin{minipage}{0.39\linewidth}
    \begin{mdframed}[roundcorner=10pt, linewidth=1.5pt]
        \centering
        \textit{Converged structure:} \\
        PBE \\
        \subfloat[\centering PBE, trial II]{{
        % \label{fig:}
        \includegraphics[width=0.80\linewidth]{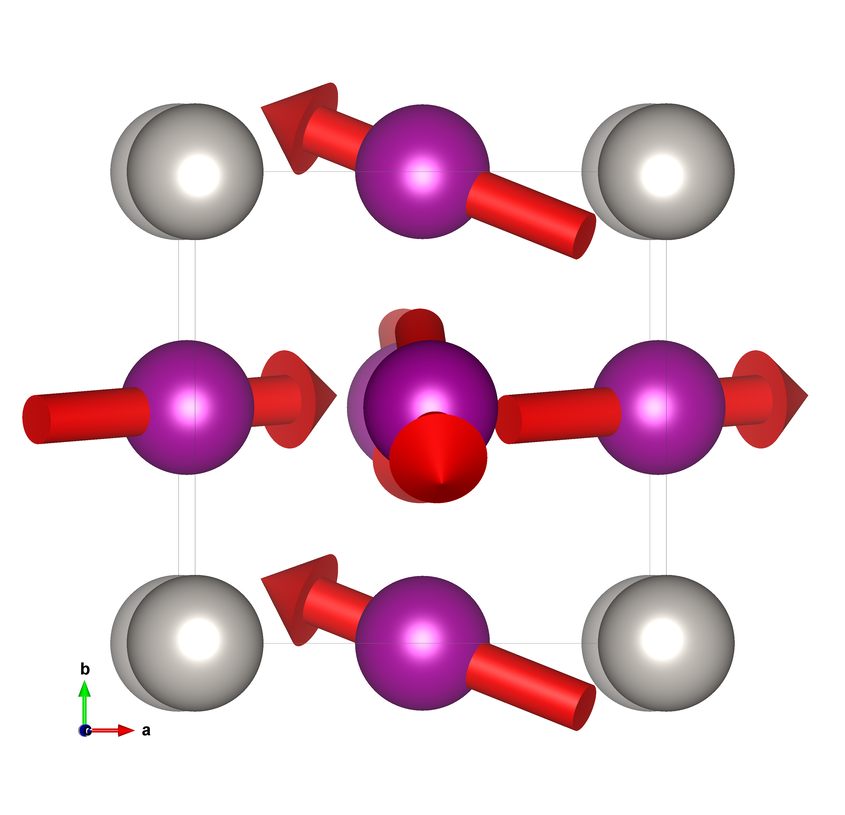}}}
    \end{mdframed}
    \end{minipage}
    \begin{minipage}{0.39\linewidth}
    \begin{mdframed}[roundcorner=10pt, linewidth=1.5pt]
        \centering
        \textit{Converged structure:} \\
        Source-free PBE \\
        \subfloat[\centering \Fsf{PBE}, trial II]{{
        %\label{fig:}
        \includegraphics[width=0.80\linewidth]{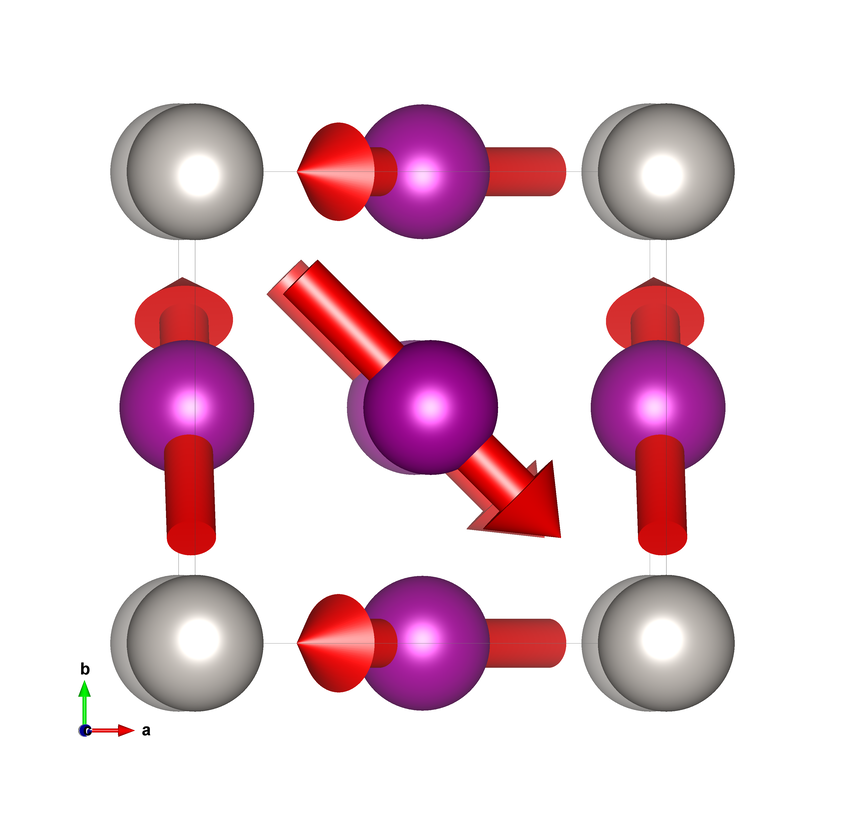}}}
    \end{mdframed}
    \end{minipage}
    %\end{mdframed}
    %%
    %\begin{mdframed}[roundcorner=10pt, linewidth=1.5pt]
    \centering
    \vspace{2ex} \\
    {\large Trial III} \\
    \vspace{2ex}
    \begin{minipage}{0.39\linewidth}
    \begin{mdframed}[roundcorner=10pt, linewidth=1.5pt]
        \centering
        \textit{Converged structure:} \\
        PBE \\
        \subfloat[\centering PBE, trial III]{{
        % \label{fig:}
        \includegraphics[width=0.80\linewidth]{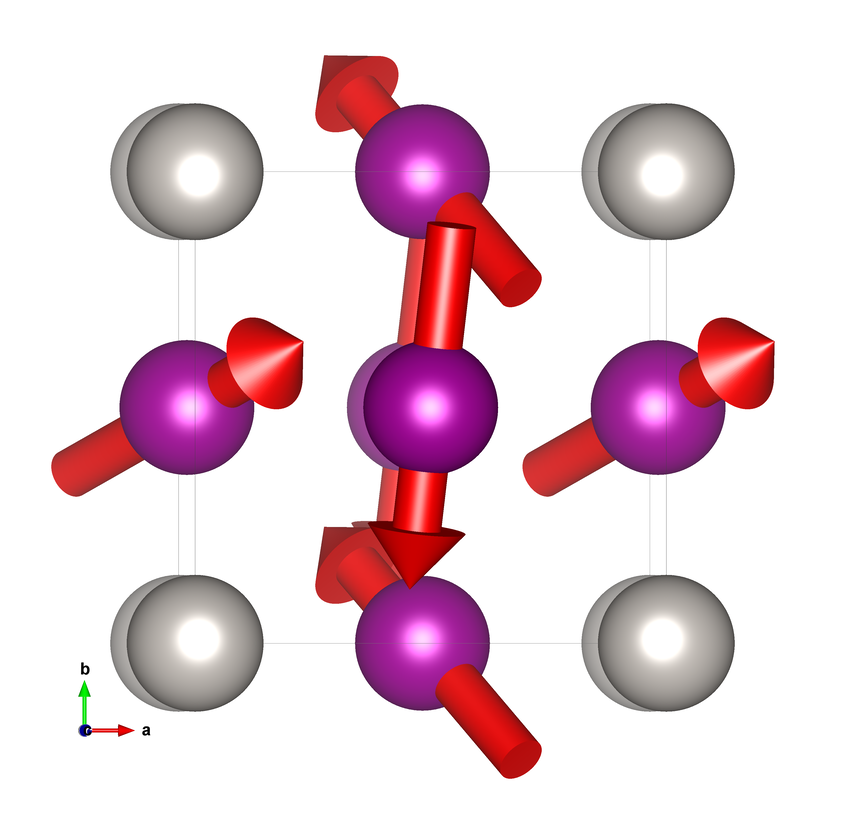}}}
    \end{mdframed}
    \end{minipage}
    \begin{minipage}{0.39\linewidth}
    \begin{mdframed}[roundcorner=10pt, linewidth=1.5pt]
        \centering
        \textit{Converged structure:} \\
        Source-free PBE \\
        \subfloat[\centering \Fsf{PBE}, trial III]{{
        %\label{fig:}
        \includegraphics[width=0.80\linewidth]{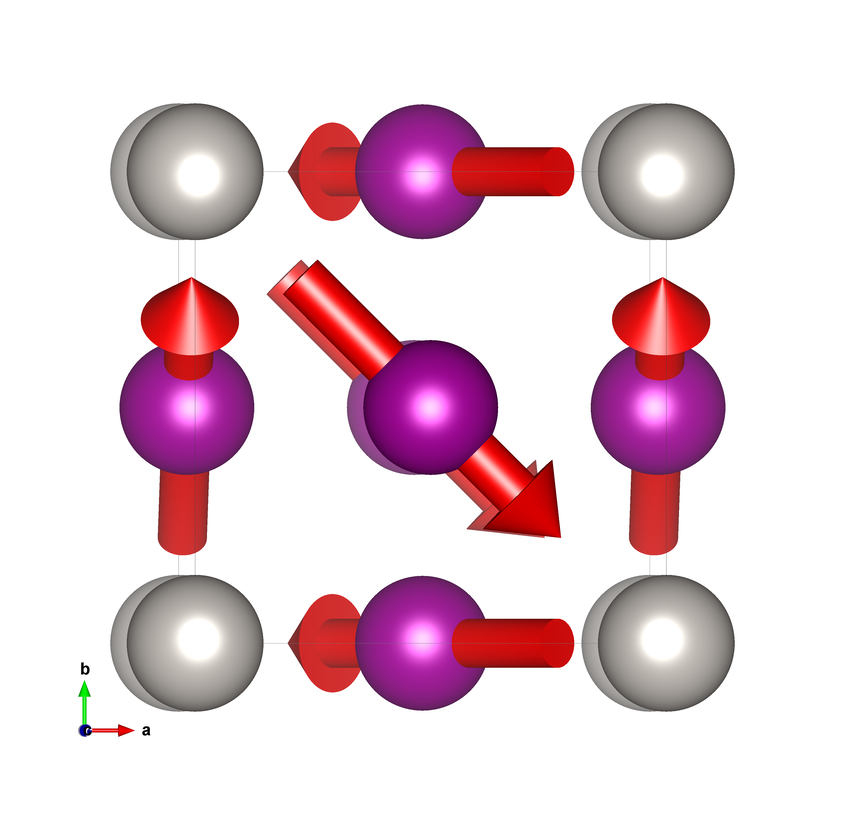}}}
    \end{mdframed}
    \end{minipage}
    %\end{mdframed}
    %%
    \caption{ \SpinPSOtoSFcaption{Mn3Pt} }
    \label{fig:Mn3Pt-SpinPSOtoSF}
\end{figure}

\begin{figure}[h]
    \centering
    %%
    %\begin{mdframed}[roundcorner=10pt, linewidth=1.5pt]
    \centering
    {\large Trial I} \\
    \vspace{2ex}
    \begin{minipage}{0.49\linewidth}
    \begin{mdframed}[roundcorner=10pt, linewidth=1.5pt]
        \centering
        \textit {Converged structure:} \\ PBE \\
        \subfloat[\centering PBE, trial I \newline \textit{viewed along [001]}]{{
        %\label{fig:}
        \includegraphics[width=0.64\linewidth]{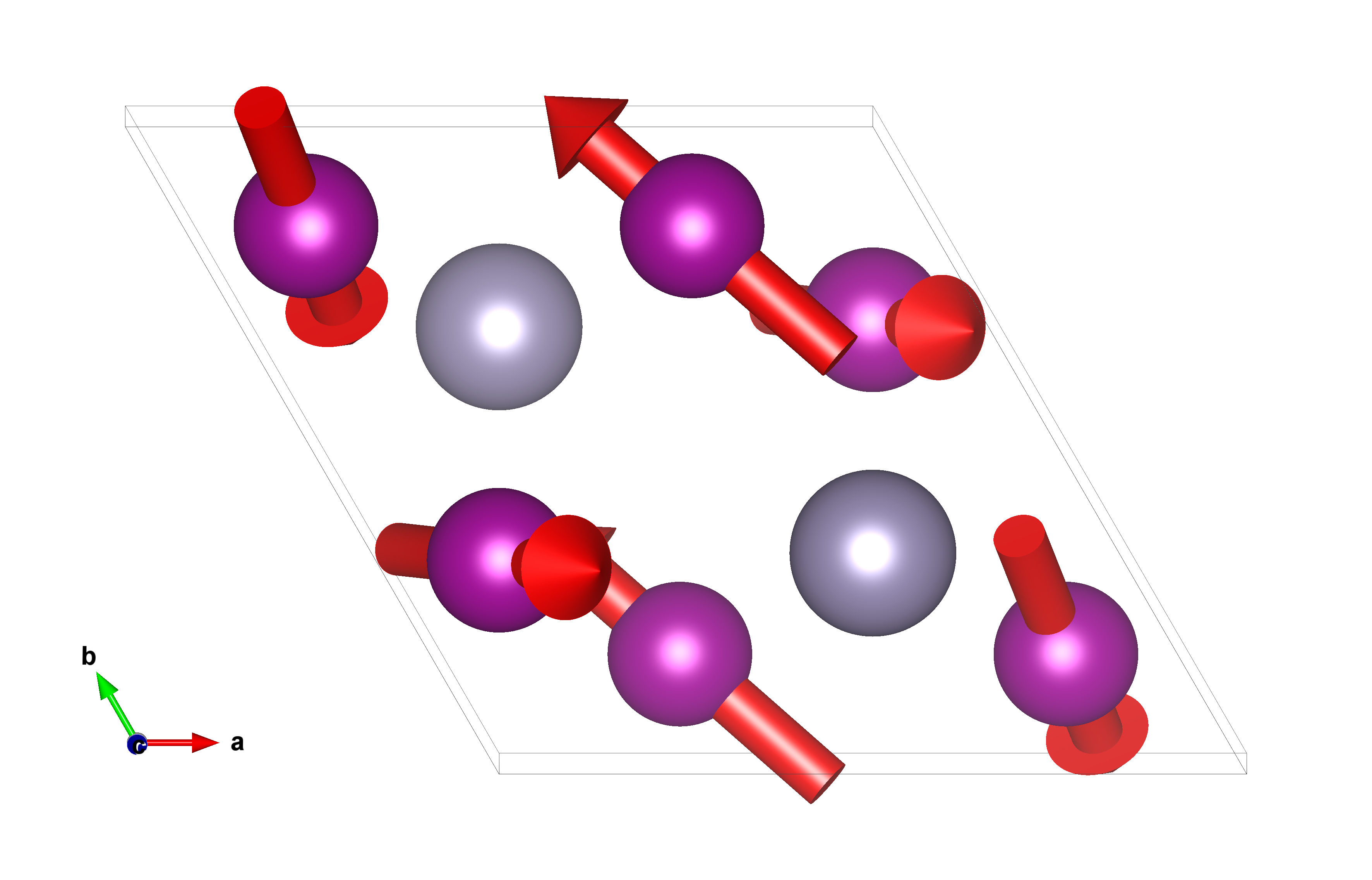}}}
        \subfloat[\centering PBE, trial I \newline \textit{viewed along [010]}]{{
        %\label{fig:}
        \includegraphics[width=0.34\linewidth]{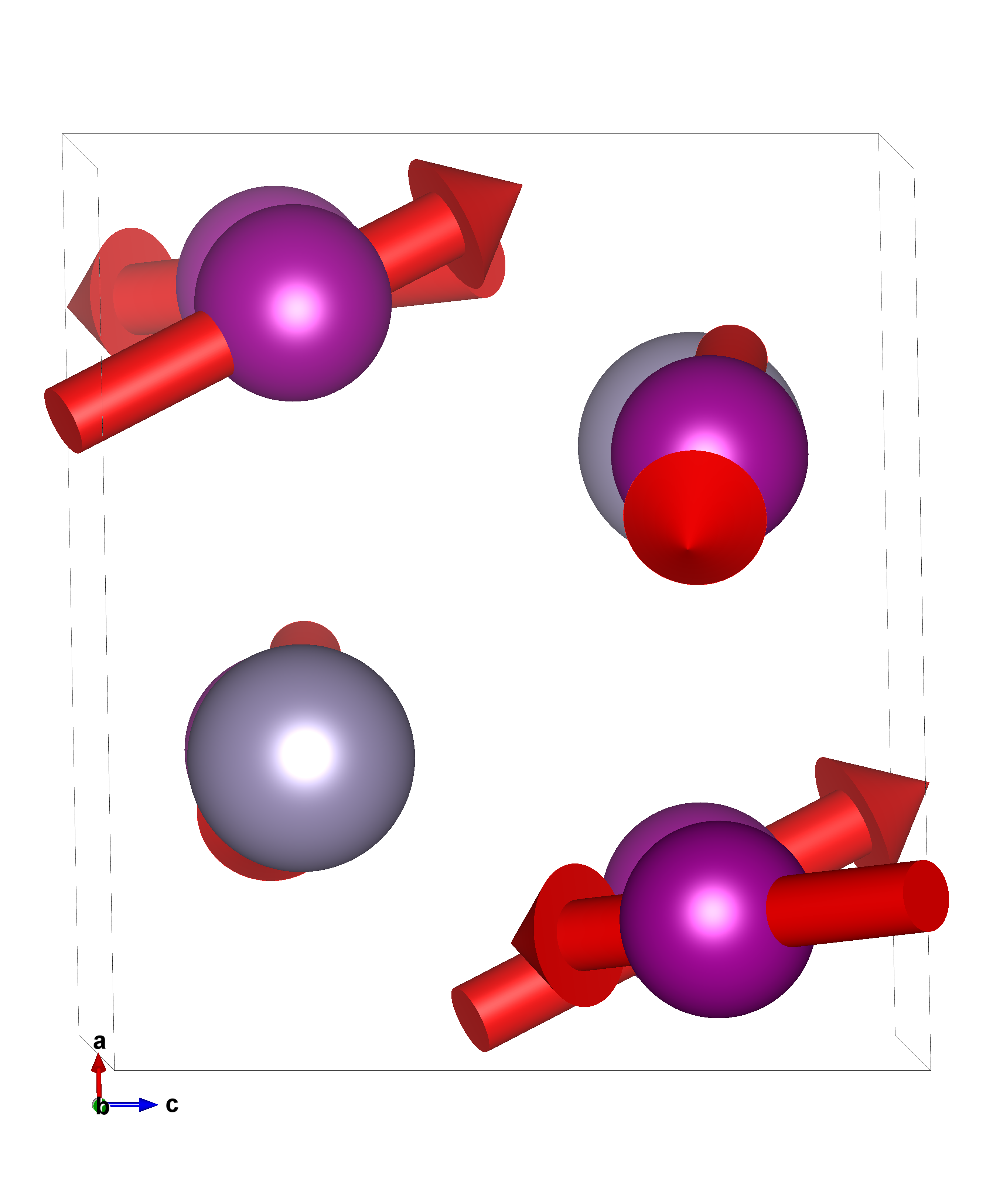}}}
    \end{mdframed}
    \end{minipage}
    \begin{minipage}{0.49\linewidth}
    \begin{mdframed}[roundcorner=10pt, linewidth=1.5pt]
        \centering
        \textit {Converged structure:} \\ Source-free PBE \\
        \subfloat[\centering PBE\textsubscript{SF}, trial I \newline \textit{viewed along [001]}]{{
        % \label{fig:}
        \includegraphics[width=0.64\linewidth]{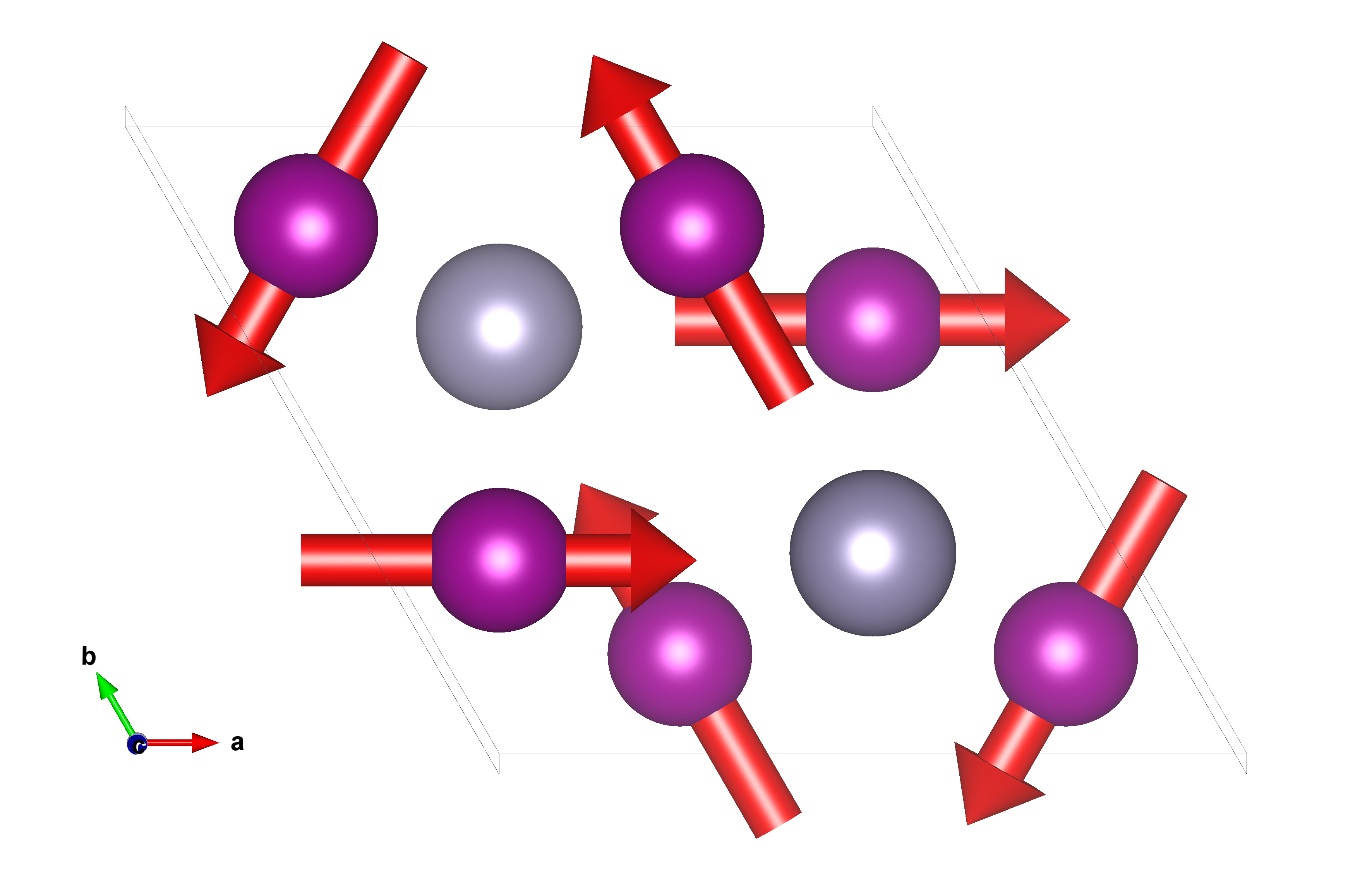}}}
        \subfloat[\centering PBE\textsubscript{SF}, trial I \newline \textit{viewed along [010]}]{{
        % \label{fig:}
        \includegraphics[width=0.34\linewidth]{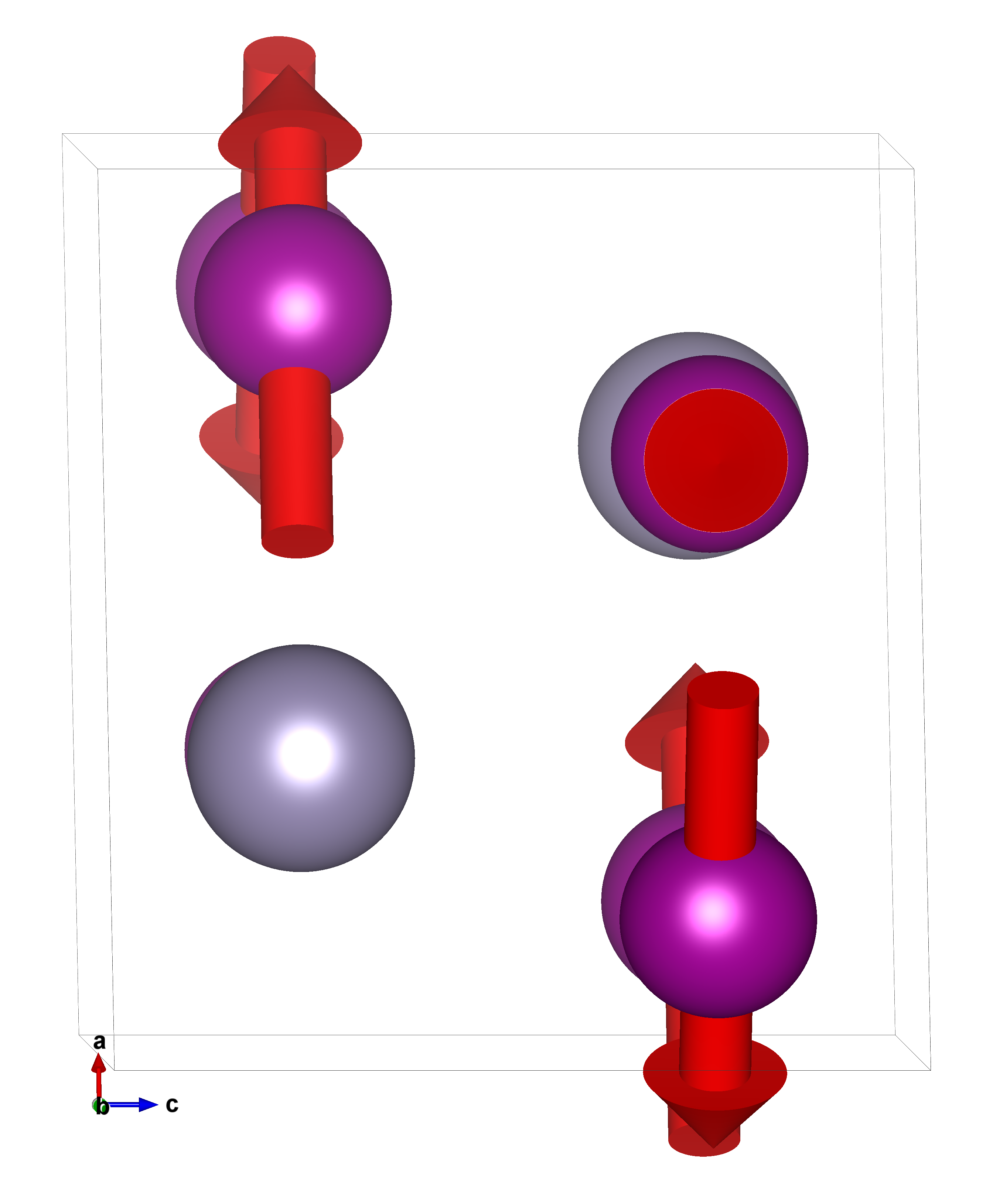}}}
    \end{mdframed}
    \end{minipage}
    %\end{mdframed}
    %%
    %\begin{mdframed}[roundcorner=10pt, linewidth=1.5pt]
    \centering
    \vspace{2ex} \\
    {\large Trial II} \\
    \vspace{2ex}
    \begin{minipage}{0.49\linewidth}
    \begin{mdframed}[roundcorner=10pt, linewidth=1.5pt]
        \centering
        \textit {Converged structure:} \\ PBE \\
        \subfloat[\centering PBE, trial I \newline \textit{viewed along [001]}]{{
        %\label{fig:}
        \includegraphics[width=0.64\linewidth]{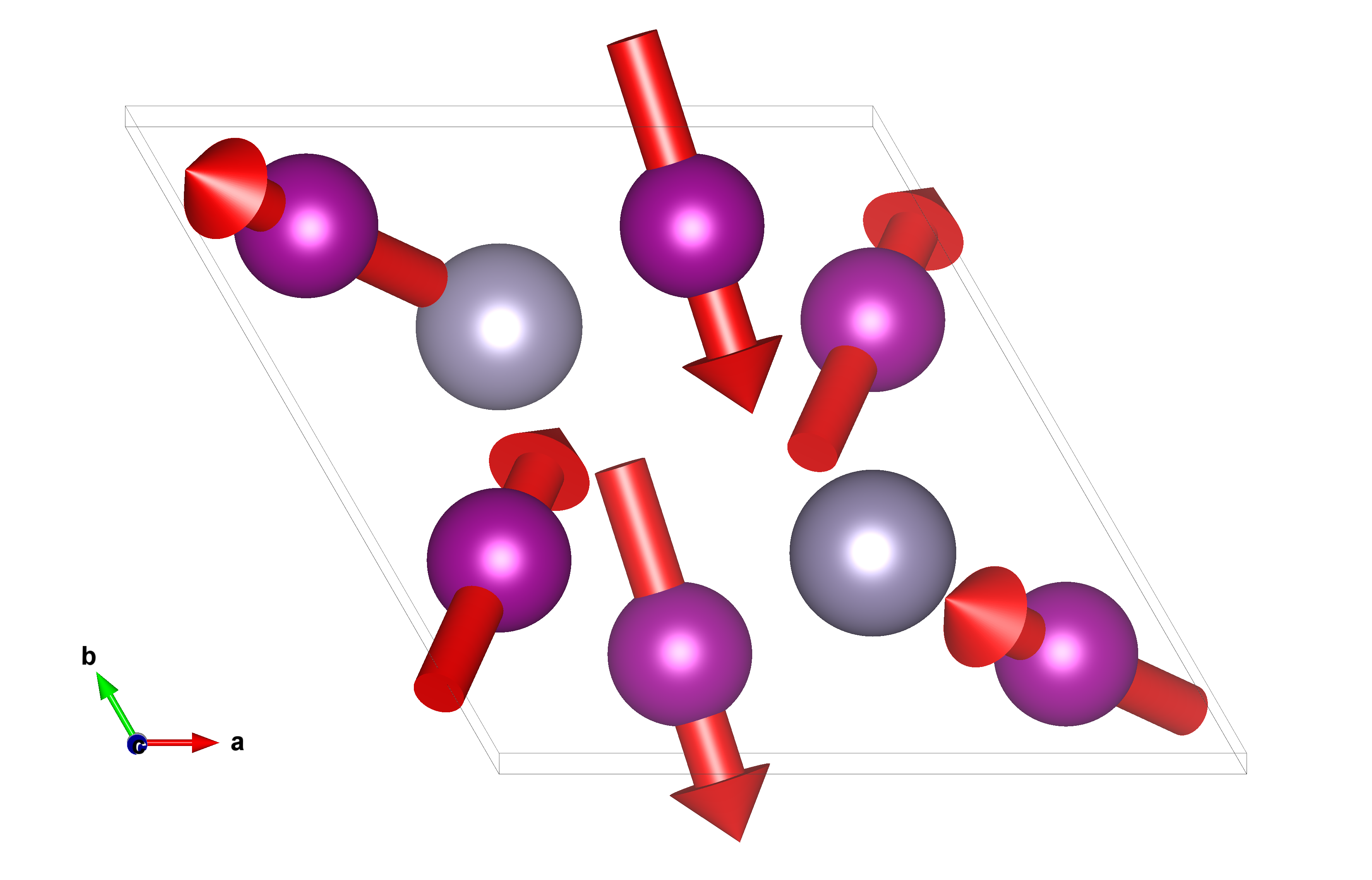}}}
        \subfloat[\centering PBE, trial I \newline \textit{viewed along [010]}]{{
        %\label{fig:}
        \includegraphics[width=0.34\linewidth]{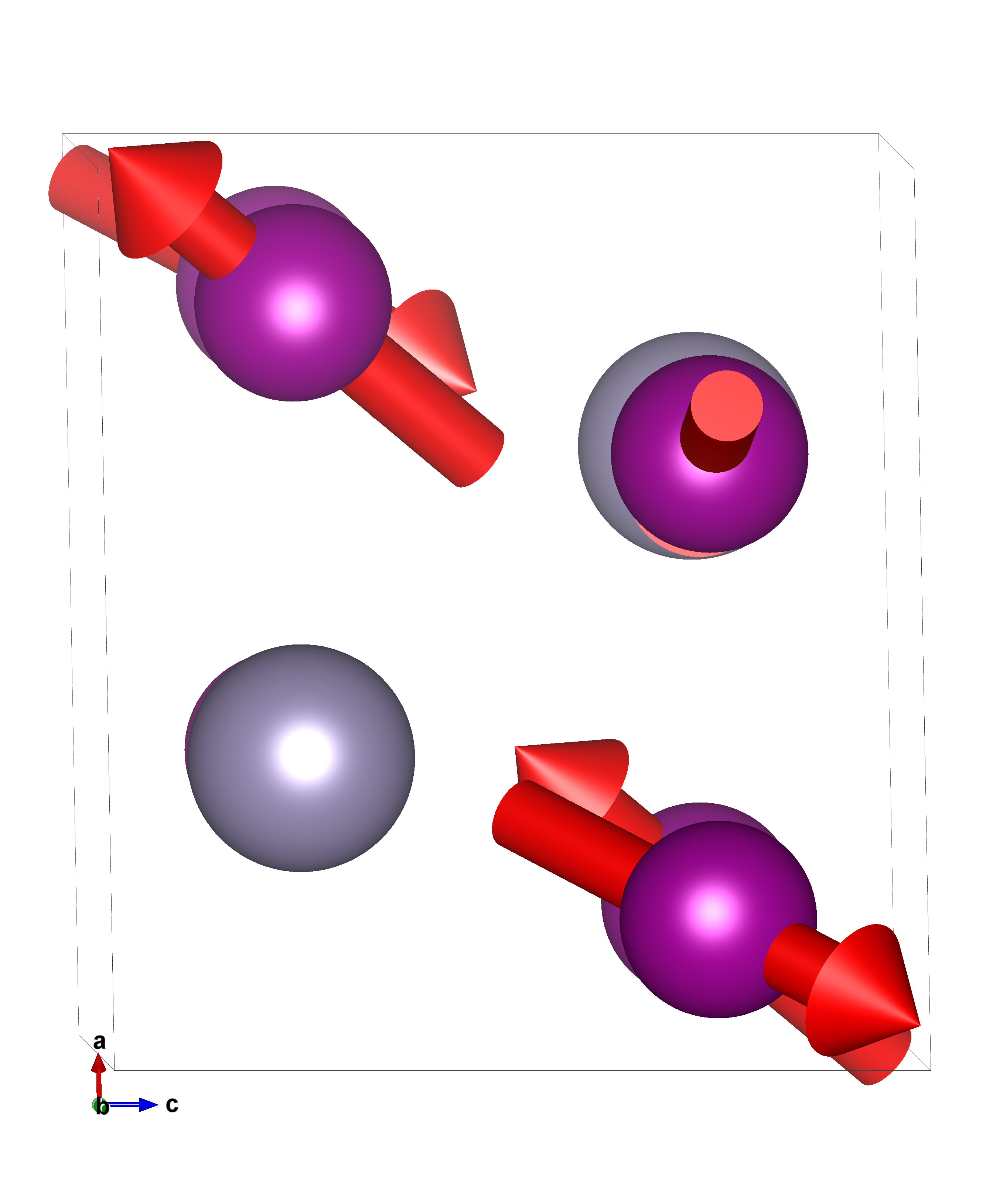}}}
    \end{mdframed}
    \end{minipage}
    \begin{minipage}{0.49\linewidth}
    \begin{mdframed}[roundcorner=10pt, linewidth=1.5pt]
        \centering
        \textit {Converged structure:} \\ Source-free PBE \\
        \subfloat[\centering PBE\textsubscript{SF}, trial I \newline \textit{viewed along [001]}]{{
        % \label{fig:}
        \includegraphics[width=0.64\linewidth]{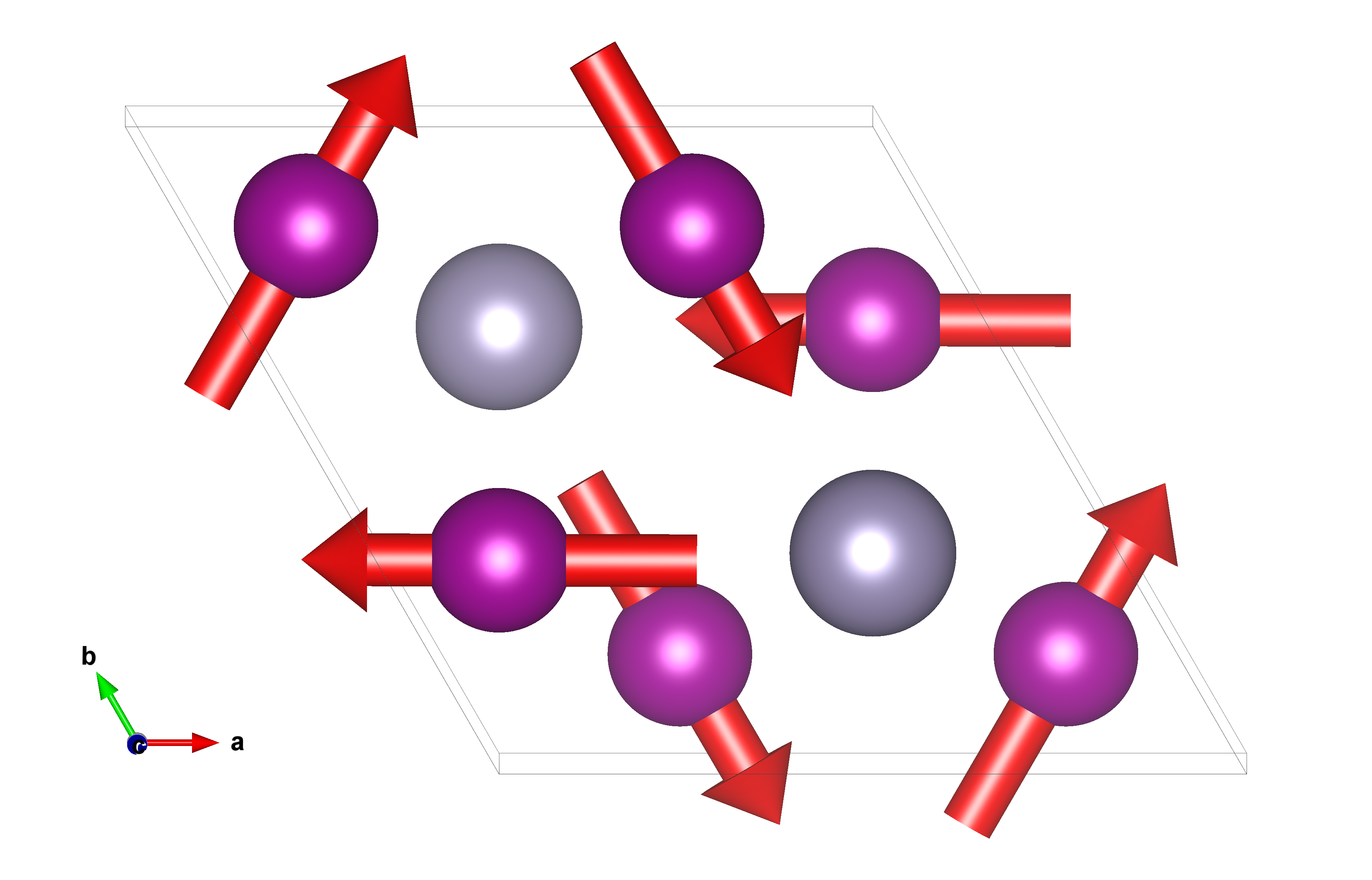}}}
        \subfloat[\centering PBE\textsubscript{SF}, trial I \newline \textit{viewed along [010]}]{{
        % \label{fig:}
        \includegraphics[width=0.34\linewidth]{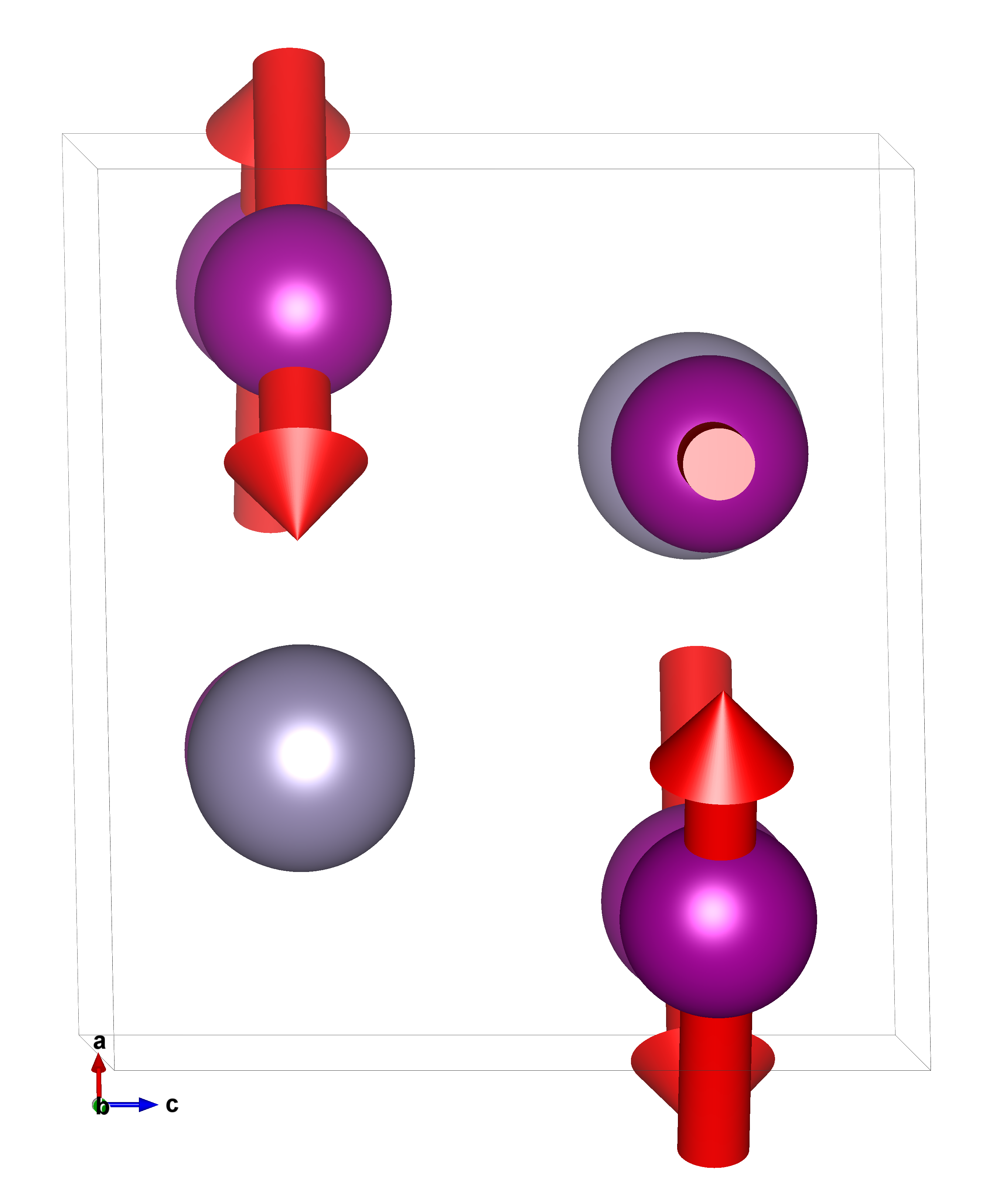}}}
    \end{mdframed}
    \end{minipage}
    %\end{mdframed}
    %%
    \caption{ \SpinPSOtoSFcaption{Mn3Sn} }
    \label{fig:Mn3Sn-SpinPSOtoSF}
\end{figure}

\begin{figure}[h]
    \centering
    %\begin{mdframed}[roundcorner=10pt, linewidth=1.5pt]
    \centering
    {\large Trial I} \\
    \vspace{2ex}
    \begin{minipage}{0.39\linewidth}
    \begin{mdframed}[roundcorner=10pt, linewidth=1.5pt]
        \centering
        \textit{Converged structure:} \\
        PBE \\
        \subfloat[\centering PBE, trial I]{{
        % \label{fig:}
        \includegraphics[width=0.98\linewidth]{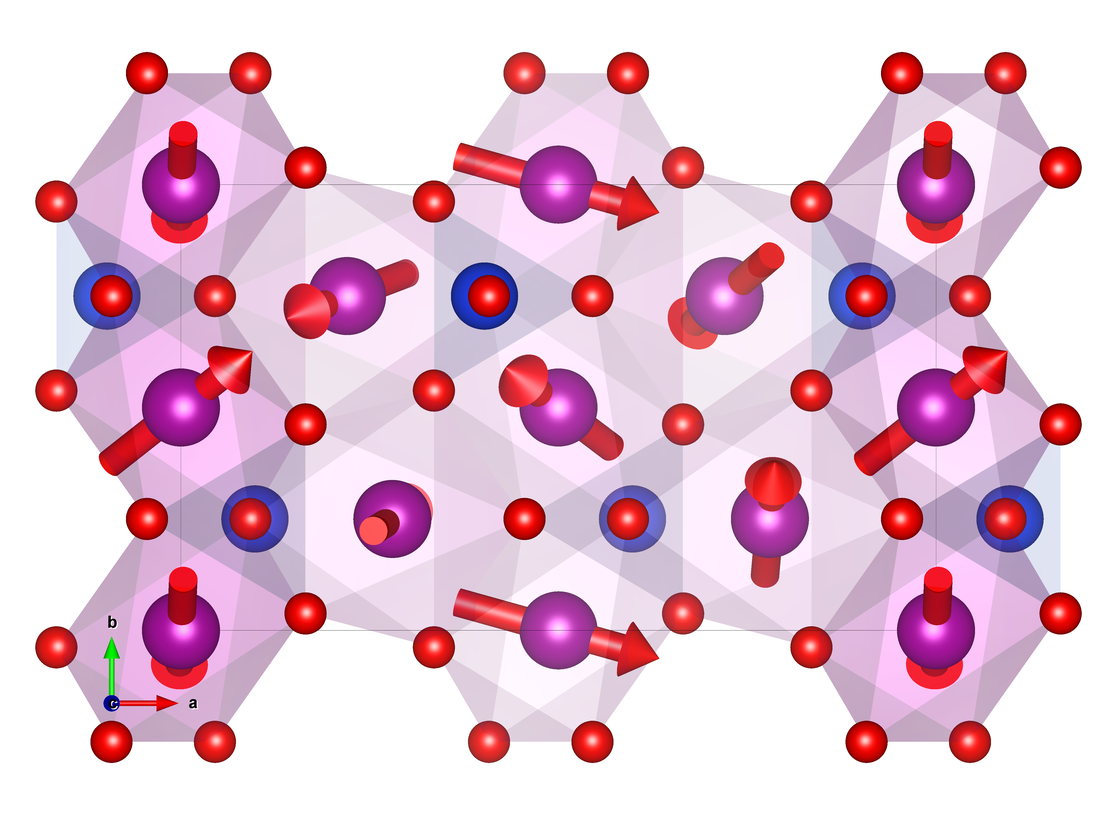}}}
    \end{mdframed}
    \end{minipage}
    \begin{minipage}{0.39\linewidth}
    \begin{mdframed}[roundcorner=10pt, linewidth=1.5pt]
        \centering
        \textit{Converged structure:} \\
        Source-free PBE \\
        \subfloat[\centering \Fsf{PBE}, trial I]{{
        %\label{fig:}
        \includegraphics[width=0.98\linewidth]{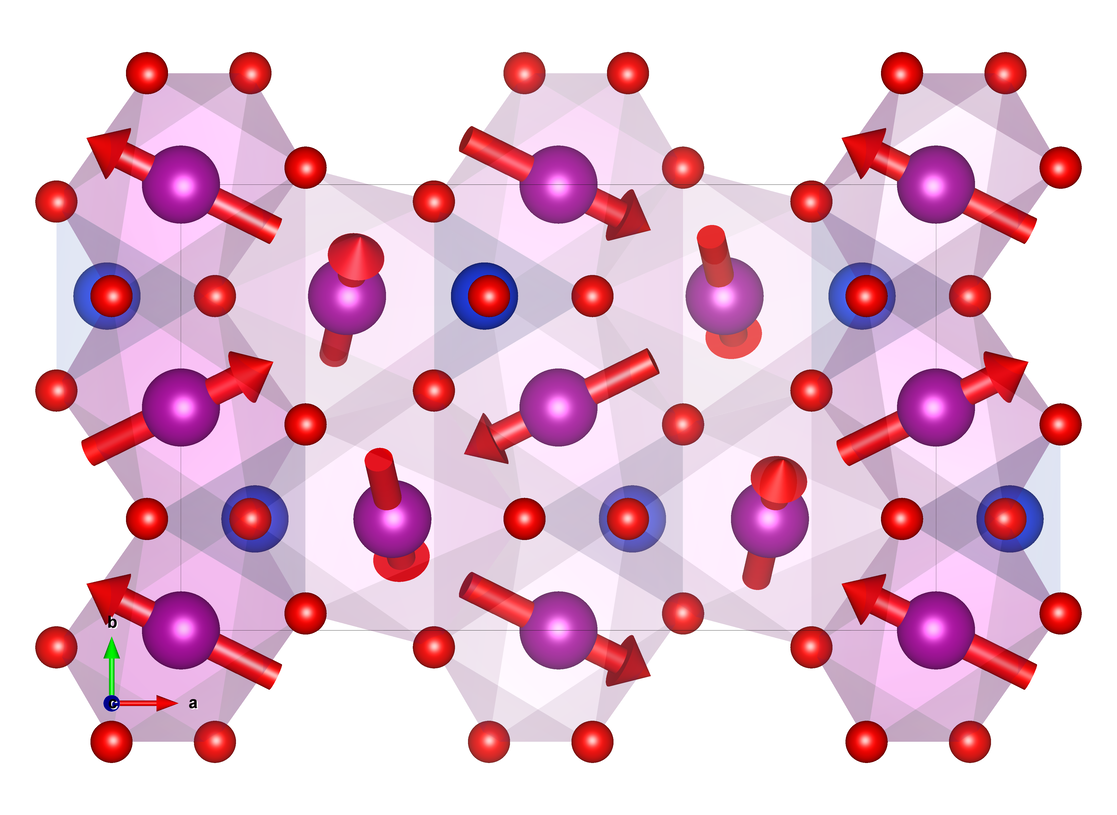}}}
    \end{mdframed}
    \end{minipage}
    %\end{mdframed}
    %%
    %\begin{mdframed}[roundcorner=10pt, linewidth=1.5pt]
    \centering
    \vspace{2ex} \\
    {\large Trial II} \\
    \vspace{2ex}
    \begin{minipage}{0.39\linewidth}
    \begin{mdframed}[roundcorner=10pt, linewidth=1.5pt]
        \centering
        \textit{Converged structure:} \\
        PBE \\
        \subfloat[\centering PBE, trial II]{{
        % \label{fig:}
        \includegraphics[width=0.98\linewidth]{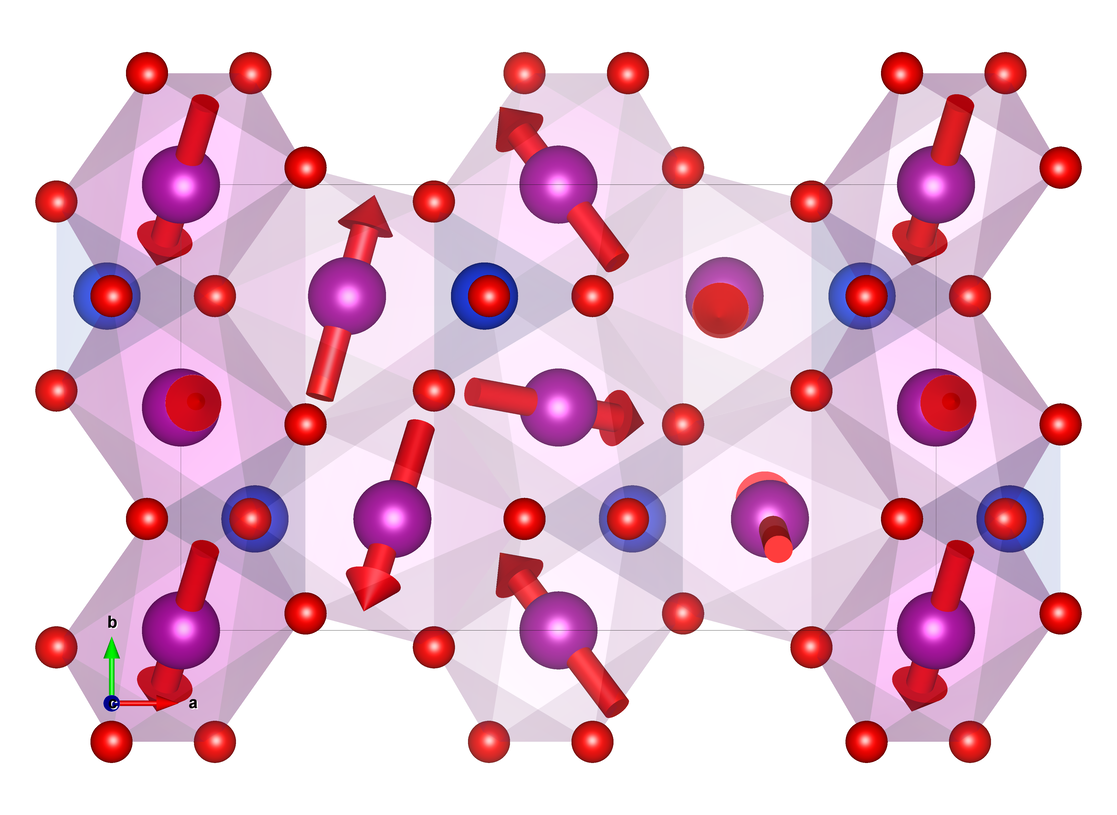}}}
    \end{mdframed}
    \end{minipage}
    \begin{minipage}{0.39\linewidth}
    \begin{mdframed}[roundcorner=10pt, linewidth=1.5pt]
        \centering
        \textit{Converged structure:} \\
        Source-free PBE \\
        \subfloat[\centering \Fsf{PBE}, trial II]{{
        %\label{fig:}
        \includegraphics[width=0.98\linewidth]{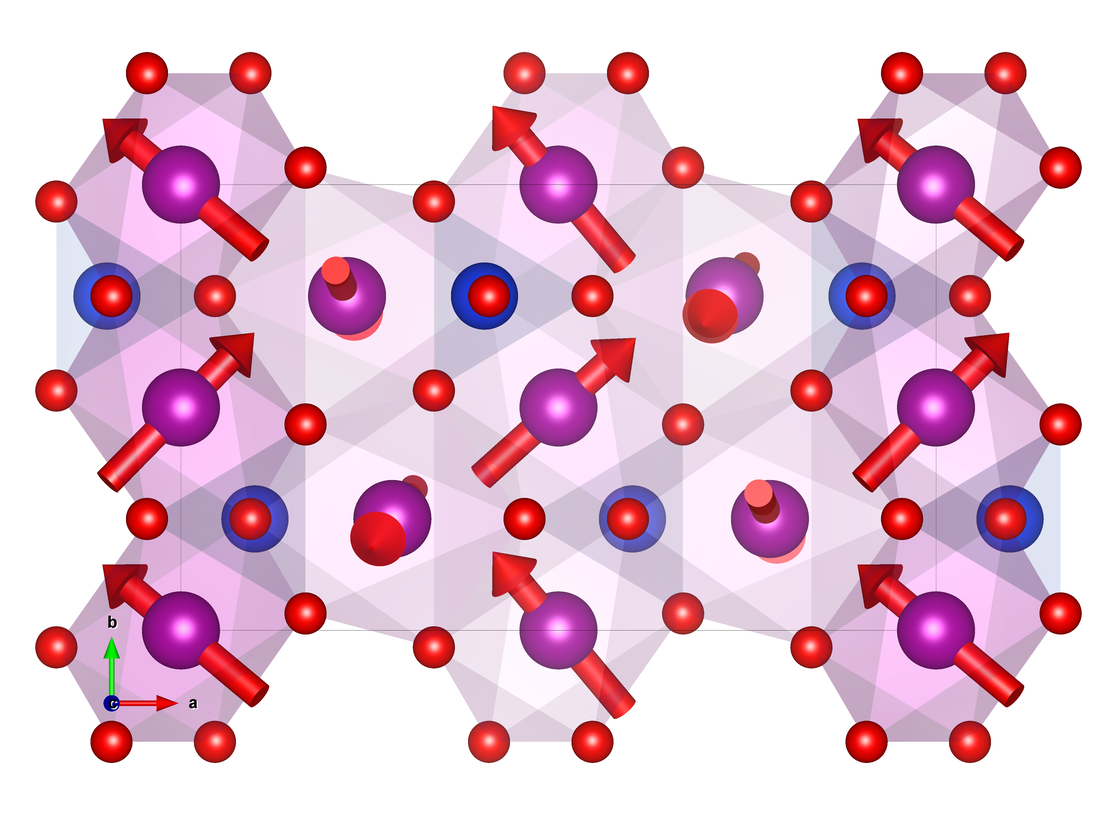}}}
    \end{mdframed}
    \end{minipage}
    %\end{mdframed}
    %%
    \caption{ \SpinPSOtoSFcaption{Mn2SiO4} }
    \label{fig:Mn2SiO4-SpinPSOtoSF}
\end{figure}

\begin{figure}[h]
    \centering
    \includegraphics[width=0.55\linewidth]{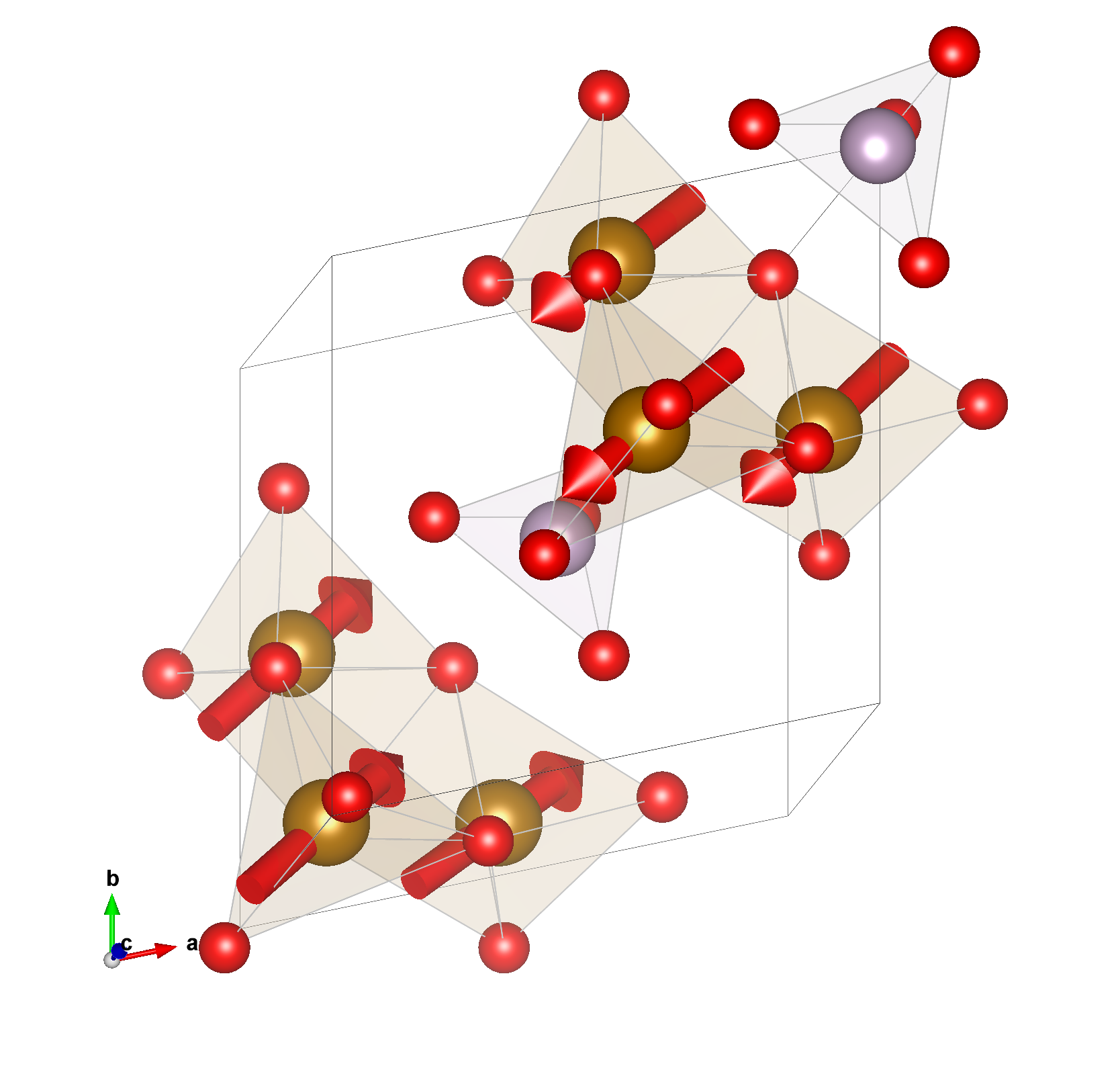} 
    \caption{Commensurate ($\bm q = \bm 0$) magnetic ground state of \ce{Fe3PO3O4}, obtained using \spso{} with PBE. This magnetic structure agrees with the experimental magnetic ground state obtained from neutron diffraction \cite{rossNanosizedHelicalMagnetic2015, tarneTuningAntiferromagneticHelical2017}. }
    \label{fig:Fe3PO7-spinpso}
\end{figure}

\end{widetext}

%%%%%%%%%%%%%%%%%%%%%%

% \nocite{*}
\bibliography{main}

%%%%%%%%%%%%%%%%%%%%%%

\end{document}